\documentclass{article} %すべて英文にしたらdvipdfmxを外してcompilerをpdfLaTeXにする
\usepackage[a4paper, top=35truemm, bottom=30truemm, left=24truemm, right=24truemm]{geometry}
\usepackage{graphicx}
\usepackage{xcolor}
\usepackage{amssymb}
\usepackage{mathrsfs}
\usepackage{tensor}
\usepackage{mathtools}
\renewcommand{\thefootnote}{\arabic{footnote}}
\usepackage{enumitem}
\usepackage{authblk}
\usepackage{pdfpages}
\usepackage{cite}
\numberwithin{equation}{section} %式番号
\usepackage{hyperref}
\hypersetup{colorlinks=true, allcolors=blue, linktocpage=true}

\DeclareMathOperator\Arccos{Arccos}

\newcommand{\ctext}[1]{\raise0.2ex\hbox{\textcircled{\scriptsize{#1}}}\,}
\newcommand{\ps}{$(+)$\,}
\newcommand{\ms}{$(-)$\,}

\renewcommand{\d}{\mathrm{d}}
\renewcommand{\Re}{\mathrm{Re}\,}
\renewcommand{\Im}{\mathrm{Im}\,}
\newcommand{\abs}[1]{|#1|}
\renewcommand{\O}[1]{\mathcal{O}\left(#1\right)}
\newcommand{\rp}{r_{(+)}}
\renewcommand{\rm}{r_{(-)}}

\newcommand{\e}{\mathrm{e}}
\newcommand{\I}{\mathrm{i}}
\newcommand{\X}{\Theta_{-\infty}}
\newcommand{\thm}{\theta_{(-)}}
\newcommand{\Q}{P} %scalar charge

\def\thefootnote{\fnsymbol{footnote}}

\begin{document}

\title{Charged multi-sheet wormhole solutions}
\author[1]{Yusuke Makita${}^{*,}$}
\author[2,3]{Keisuke Izumi}
\author[4]{Daisuke Yoshida}

\affil[1]{Graduate School of Science, Nagoya University, Nagoya 464-8602, Japan}
\affil[2]{Department of Mathematics, Nagoya University, Nagoya 464-8602, Japan}
\affil[3]{Kobayashi-Maskawa Institute, Nagoya University, Nagoya 464-8602, Japan}
\affil[4]{Institute for Advanced Research, Nagoya University, Nagoya 464-8602, Japan}
\date{}
\maketitle
\vspace{-10mm}
\begin{tabbing}
    \hspace{10mm} E-mail: \ \= \href{mailto:makita.yusuke.c2@s.mail.nagoya-u.ac.jp}{makita.yusuke.c2@s.mail.nagoya-u.ac.jp}, \\ \>
    \href{mailto:izumi@math.nagoya-u.ac.jp}{izumi@math.nagoya-u.ac.jp}, \\ \>
    \href{mailto:yoshida.daisuke.k9@f.mail.nagoya-u.ac.jp}{yoshida.daisuke.k9@f.mail.nagoya-u.ac.jp} \\
     \hspace{10mm} ${}^{*}$ Author to whom any correspondence should be addressed.
\end{tabbing}

\renewcommand\thefootnote{\arabic{footnote}}

\vspace{5mm}
\begin{abstract}
We construct charged wormhole solutions with an even number of asymptotically flat regions in the four-dimensional Einstein-Maxwell-massless phantom scalar system via the Harrison transformation.
The solutions are characterized by five parameters: the mass~$M$, the electric charge~$Q_\mathrm{e}$, the magnetic charge~$Q_\mathrm{m}$, the scalar charge~$\Q$ and the number of sheets $2n$. The regularity condition then determines the throat radius.
Although the Harrison transformation directly generates the solutions only in the parameter region $Q_{\mathrm{e}}^2 + Q_{\mathrm{m}}^2 < M^2$, 
we show that regular solutions exist in a wider parameter region beyond this bound.
In addition, we introduce a spheroidal coordinate system that covers one complete asymptotically flat region and its adjacent ones, and allows the solution to be expressed in a simple form.
\end{abstract}

%\tableofcontents

%\newpage
\section{Introduction}
Exact solutions in general relativity provide concrete laboratories for exploring the nonlinear and global properties of spacetime. Since the Einstein equations are highly nonlinear partial differential equations, many solutions have been constructed by imposing symmetry assumptions. 

The Schwarzschild solution is one of the simplest solutions of the vacuum Einstein equations, obtained under the assumption of the static and spherical symmetry.
It describes a black hole spacetime for a positive Arnowitt–Deser–Misner (ADM) mass $M > 0$.
The Reissner-Nordstr\"om solution is its charged counterpart in the Einstein-Maxwell system.
A characteristic feature of the Reissner-Nordstr\"om solution is that its causal structure depends crucially on the mass-charge ratio: it describes a charged black hole for $M \geq Q$, while the horizons disappear and a naked singularity appears for $Q > M$, where $Q = \sqrt{Q_{\mathrm{e}}^2 + Q_{\mathrm{m}}^2}$, with $Q_{\mathrm{e}}$ and $Q_{\mathrm{m}}$ denoting the electric charge and magnetic charge in geometric units.
Thus, electromagnetic charge can qualitatively change the global structure of spacetime.
This illustrates the importance of studying charged exact solutions as explicit examples of how matter fields and charges affect spacetime geometry.

For stationary and axisymmetric spacetimes, the Einstein-Maxwell equations admit a remarkable reformulation due to Ernst~\cite{Ernst1968,Ernst1968b}.
Ernst pointed out the existence of an integrable structure within the Einstein-Maxwell system under an ansatz of stationary and axial symmetry.
The set of reduced field equations, called \textit{the Ernst equations}, consists of two complex potentials: the Ernst potential $\mathcal{E}$ associated with geometric components, and the electromagnetic potential $\Phi$.
Building on Ernst's formulation, Harrison introduced a solution-generating transformation, now known as the Harrison transformation~\cite{Harrison1968}.
This transformation generates a one-complex-parameter family of solutions to the Einstein-Maxwell equations from a given seed solution.
In particular, it can be used to construct a charged solution from a vacuum solution.
For example, the Reissner-Nordstr\"om solution can be generated from the Schwarzschild solution, 
as reviewed in Sec.~\ref{schwarzschild}.

Following the development of the Ernst formalism,
Belinski and Zakharov established a solution generating technique based on the inverse scattering method, called \textit{the gravitational soliton formalism}~\cite{Belinski1978, Belinski1979}.
Its ansatz is the same as that of the Ernst equation for the vacuum system, and it reduces a non-linear differential problem to a kind of linear problem.
Although it has been successfully applied to the vacuum Einstein equations, its extension to the electrovacuum case is more involved, and only a few attempts have been made in this direction~\cite{Belinski:2001ph}.

The Ernst formalism, the gravitational soliton formalism, and their solution-generating technique, have been widely
used to construct exact solutions.
An example is the class of the Tomimatsu-Sato solutions obtained by solving the Ernst equations, which generalize the Kerr solution but contain naked singularities~\cite{Tomimatsu:1972zz}.
Another example is the Ernst-Wild solutions describing black holes immersed in a magnetic field, which are generated via the Harrison transformation from static black hole seeds~\cite{Ernst:1976mzr} and stationary seeds~\cite{Ernst:1976bsr}.

The Ernst formalism and the gravitational soliton formalism can still be applied in the presence of a massless scalar field.
This observation allows us to use the solution-generating techniques to explore a wider class of exact solutions with nontrivial spacetime structures.
In particular, when the scalar field is a so-called phantom scalar field, it can support wormhole geometries and lead to spacetime structures that are absent in the ordinary electrovacuum system.
Topological censorship theorem~\cite{Friedman:1993ty,Friedman:2006tea,Graham:2007va} suggests that the wormhole structure which connects two or more distinct asymptotic regions requires violation of the achronal averaged null energy condition, such as a phantom scalar field.

Exact wormhole solutions in the system with a phantom scalar field have so far been studied mainly in static and spherically symmetric assumptions.
Actually, the Ellis drainhole solution~\cite{Ellis1973,Bronnikov}, which is now known as Ellis wormhole, connecting two distinct spatial infinities is such an example.
The Ellis wormhole was later extended to include a nonvanishing ADM mass in Ref.~\cite{Armendariz-Picon:2002gjc},
and an electric charge in Ref.~\cite{Turimov:2025zxy}, although spherical symmetry was still assumed.

In our previous work with a collaborator \cite{Makita:2025bao}, we investigated exact wormhole solutions in the Einstein-phantom scalar system under the assumption of static axisymmetry.
There, we found exact solutions that connect an even number of asymptotic flat regions\footnote{
While solutions called multi-wormholes were also discussed by Gibbons and Volkov~\cite{Gibbons:2017jzk}, our multi-sheet structure, which is characterized by multiple asymptotic regions branching from a single throat, were first introduced in our paper \cite{Makita:2025bao}, where the explicit regularity conditions were also established.
}.
An important observation is that the wormhole solution can be formulated within the Weyl formalism, which can be regarded as a static subclass of the Ernst formalism.
This suggests that solution-generating transformations in the Ernst formalism can be applied to these wormhole solutions.
In particular, the Harrison transformation provides a natural way to construct charged generalizations of the multi-sheet wormholes.

In this paper, we construct charged generalizations of multi-sheet wormholes by applying the Harrison transformation to the solutions obtained in our previous work.
We study their regularity and show that the family of regular solutions extends beyond the parameter region directly generated by the Harrison transformation.

This paper is organized as follows.
In Sec.~\ref{generate}, we review the Ernst formalism and the Harrison transformation with an example of application.
In Sec.~\ref{charged}, we construct the charged multi-sheet wormhole solutions via the Harrison transformation and analyze their regularity conditions.
In Sec.~\ref{extension}, we extend the family of regular solutions beyond the parameter region 
directly generated by the Harrison transformation.
In Sec.~\ref{spheroid}, we introduce a coordinate system adapted to a pair of asymptotically flat regions, in which the solutions take a simple form.
Finally, in Sec.~\ref{summary}, we provide a brief summary of this article.
In Appendix~\ref{analyses}, we give details of the regularity analyses performed in Sec.~\ref{charged}.
Throughout this paper, we use geometric units where the speed of light, the Newton constant and the Coulomb constant are set to unity.

\section{Solution generating technique} \label{generate}
In this section, we review a solution-generating technique, the Harrison transformation within the Ernst formalism, as a preparation for subsequent sections. 
We extend the formalism to include a general matter field satisfying the condition given in Eq.~\eqref{matter1} below. 
A massless scalar field, which can be either a canonical or a phantom scalar, satisfies this condition.

In the next section, we apply this formalism to construct static, electromagnetized wormhole solutions.
While the applications discussed in this paper are restricted to static cases, 
we introduce the formalism here in its general form in the presence of rotation.

\subsection{The Ernst formalism} \label{ernst}
The Ernst formalism provides a simplification by imposing certain symmetries, 
typically those of stationarity and axisymmetry. 
This subsection provides a brief review of this formalism.
For further details, see, \textit{e.g.},
\cite{Vigano:2022hrg, Griffiths_Podolsky_2009, Stephani:2003tm}.

Firstly, we assume a four-dimensional spacetime possessing two commuting Killing vectors. 
One is typically identified as the time translation vector, $\partial_t$, 
while the other corresponds to the rotational Killing vector, $\partial_\phi$. 
Under these symmetries, the spacetime can be decomposed into two sectors;
the Killing orbits spanned by $(t, \phi)$ and 
the remaining two-dimensional subspace described by the cylindrical coordinates.
We focus on the case where $(\rho, z)$ coordinates are the Weyl canonical coordinates\footnote{
One can introduce the Weyl canonical coordinates ($\rho, z$) when $\sqrt{-\det g_{ab}}$ is a harmonic function, where $g_{ab}$ denotes the $(t, \phi)$-block of the metric.
One of the coordinates, $\rho$, is identified with $\sqrt{-\det g_{ab}}$.
The other coordinate, $z$, is defined as the harmonic conjugate of $\rho$.
} $(\rho, z)$.
Consequently, the line element can be expressed in terms of three functions of the coordinates, known as \textit{the Weyl-Papapetrou form}, as follows:
\begin{align}
    \d s^2 &= -f(\rho, z)\left(\d t-\omega(\rho, z)\d\phi\right)^2 + \frac{1}{f(\rho, z)}\left[\e^{2\gamma(\rho, z)}\left(\d\rho^2+\d z^2\right)+\rho^2\d\phi^2\right].
    \label{WP}
\end{align}

We also impose the stationarity and axisymmetry on the gauge potential by choosing a suitable gauge. 
In this paper, we assume the following form:
\begin{align}
A_\mu \d x^\mu &= A_t(\rho, z) \d t + A_\phi(\rho, z) \d \phi.
\end{align}
It should be noted that, in general, the Ernst formalism does not preclude the existence of $\rho$ and $z$ components in the Maxwell potential.
However, we assume their absence here for simplicity. 
In the limit where the electromagnetic field vanishes, the system reduces to the vacuum model investigated in our previous work \cite{Makita:2025bao}. 
The present study extends that work by incorporating the gauge potential into the Ernst formalism.

The Ernst formalism reduces the Einstein and Maxwell equations to a system of differential equations in three-dimensional Euclidean space $\mathbb{R}^3$ with axial symmetry. 
The corresponding line element is given by
\begin{align}
 \d l^2= \d \rho^2 + \d z^2 + \rho^2 \d \phi^2.
\end{align}
For later convenience, we define the scalar and cross products for any vectors $A=A_\rho \d\rho+ A_z \d z+ A_\phi \d \phi$ and $B=B_\rho \d\rho+ B_z \d z+ B_\phi \d \phi$ in $\mathbb{R}^3$ as
\begin{align}
    A\cdot B &= A_\rho B_\rho + A_zB_z + \frac{1}{\rho^2}A_\phi B_\phi, \\
    A\times B &= \frac{1}{\rho}\left(A_zB_\phi - A_\phi B_z\right)\hat\rho + \frac{1}{\rho}\left(A_\phi B_\rho - A_\rho B_\phi\right)\hat z + \rho\left(A_\rho B_z - A_zB_\rho\right)\hat\phi.
\end{align}
where $\hat\rho, \hat z$, and $\hat\phi$ are the unit basis vectors in the $\partial_\rho$, $\partial_z$, and $\partial_\phi$ directions, respectively. 
Note that we have chosen the orientation of the basis as $(\hat \rho, \hat z, \hat \phi)$.
Let $f$ be a scalar field and $A$ be a vector field on $\mathbb{R}^3$ whose components depend only on $\rho$ and $z$. The operations of the covariant derivative $D$ are then defined as follows:
\begin{align}
    D f &= \left(\partial_\rho f\right)\hat\rho + \left(\partial_zf\right)\hat z, \\
    D \cdot A &= \frac{1}{\rho}A_\rho + \partial_\rho A_\rho + \partial_z A_z, \\
    D^2 f &= \frac{1}{\rho}\partial_\rho f + \partial^2_\rho f + \partial^2_z f, \\
    D \times A &= \left(\partial_zA_\phi\right)\hat\rho - \left(\frac{1}{\rho}A_\phi + \partial_\rho A_\phi \right)\hat z + \left(\partial_\rho A_z - \partial_zA_\rho\right)\hat\phi, \\
    \hat\phi\times D f &= -\left(\partial_zf\right)\hat\rho + \left(\partial_\rho f\right)\hat z.
\end{align}

Throughout this paper, we consider the Einstein-Maxwell system with an additional matter field, whose action is given in the form,
\begin{align}
    S\left[g_{\mu\nu}, A_\mu\right] \coloneqq \int\left[ \frac{1}{2\kappa} \left( R-F_{\mu\nu}F^{\mu\nu} \right) + \mathcal{L}^{(\mathrm{m})}
    \right]\sqrt{-g}\d^4x,
\end{align}
where $\kappa = 8 \pi$ in our unit.
The equations of motion derived from this action are summarized as 
\begin{align}
R_{\mu\nu} - \frac{1}{2} R g_{\mu\nu} &= \kappa T^{(\mathrm{EM})}_{\mu\nu} + \kappa T_{\mu\nu}^{(\mathrm{m})}, \\
\nabla_{\mu} F^{\mu\nu} &= 0, \label{Maxwell eq}
\end{align}
as well as the equations of motion for matter field. 
The energy momentum tensor of the electromagnetic field is given by
\begin{align}
\kappa T_{\mu\nu}^{(\text{EM})} \coloneqq 2\left(F_{\mu \rho} F_\nu{}^\rho-\frac{1}{4}F_{\rho\sigma}F^{\rho \sigma}g_{\mu\nu}\right)
,
\end{align}
and that of the matter field is defined by
\begin{align}
    T_{\mu\nu}^{(\mathrm{m})} \coloneqq -\frac{2}{\sqrt{-g}}\frac{\delta\left(\sqrt{-g}\mathcal{L}^{(\mathrm{m})}\right)}{\delta g^{\mu\nu}} .
\end{align}

The Einstein equations can be expressed as 
\begin{align}
    R_{\mu\nu} = \kappa T^{(\mathrm{EM})}_{\mu\nu} + S^{(\mathrm{m})}_{\mu\nu}.
\end{align}
where
\begin{align}
S_{\mu\nu}^{(\mathrm{m})} \coloneqq \kappa \left(T_{\mu\nu}^{(\mathrm{m})} - \frac{1}{2} T^{(\mathrm{m})} g_{\mu\nu}\right).
\end{align}
Let us consider the case where $S_{\mu\nu}^{(\mathrm{m})}$ has non-zero components 
only in the $(\rho,z)$-subspace.
This implies 
\begin{align}
    S^{(\mathrm{m})}_{a b} = 0,
    \quad S^{(\mathrm{m})}_{a i} = 0, \label{matter1}
\end{align}
where the alphabet indices, $a, b, \cdots$, denote the coordinates for the symmetries, either $t$ or $\phi$, while the second part of them, $i, j, \cdots$, denote the coordinates in the two-dimensional plane, either $\rho$ or $z$.
Under this assumption, the Einstein equations can be expressed as
\begin{align}
R_{a b} &= \kappa T^{(\mathrm{EM})}_{ab}, \label{Rab =}\\
R_{a i} &= \kappa T^{(\mathrm{EM})}_{a i},
\label{Rai =}
\end{align}
and
\begin{align}
    R_{ij} &= \kappa T^{(\mathrm{EM})}_{ij} + S^{(\mathrm{m})}_{ij}.
    \label{einstein1}
\end{align}
In this setup, the matter field contributes to the Einstein equations only within the $(i, j)$-block. 
This specific form of the matter source ensures that the $(a, b)$ and $(a, i)$-components of the field equations remain identical to those in the electromagnetic vacuum case, 
which is a crucial requirement for the application of the Ernst formalism.
 
Ernst showed that Eqs. \eqref{Rab =}, as well as the Maxwell equations \eqref{Maxwell eq} can be reduced to a pair of two \textit{complex} potential equations, which are called \textit{the Ernst equations}\cite{Ernst1968, Ernst1968b},
\begin{align}
    \left\{
    \begin{aligned}
        \left(\Re\mathcal{E}+\abs{\Phi}^2\right)D^2\mathcal{E} &= D\mathcal{E}\cdot\left(D\mathcal{E}+2\Phi^*D\Phi\right), \\
        \left(\Re\mathcal{E}+\abs{\Phi}^2\right)D^2\Phi &= D\Phi\cdot\left(D\mathcal{E}+2\Phi^*D\Phi\right).
    \end{aligned}
    \right. \label{ernsteq}
\end{align}
Two complex scalar functions $\mathcal{E}$ and $\Phi$, called \textit{the Ernst} and \textit{electromagnetic  potentials} respectively, are defined as
\begin{align}
    \mathcal{E} &\coloneqq f - \abs{\Phi}^2 + \I\tilde\omega, \label{erpot3} \\
    \Phi &\coloneqq A_t + \I\tilde A_\phi, \label{Phipot5}
\end{align}
where $\I$ denotes the imaginary unit and the quantities with tilde symbol, $\tilde\omega$ and $\tilde A_\phi$, are \textit{twisted potentials} generated from the original variables through the relations,
\begin{align}
    \hat\phi\times D \tilde\omega &\coloneqq \frac{1}{\rho}f^2 D \omega - 2\hat\phi\times\Im\left(\Phi^*D \Phi\right), \label{erpot2} \\
    \hat\phi\times D \tilde A_\phi &\coloneqq -\frac{1}{\rho}f\left(\omega D A_t+ D  A_\phi\right). \label{erpot1}
\end{align}
Once a solution $(\mathcal{E}, \Phi)$ to the Ernst equations~\eqref{ernsteq} is obtained, one can reconstruct the metric functions $f, \omega$ and the Maxwell potentials $A_t, A_\phi$ via Eqs.~\eqref{erpot3}\nobreak-\nobreak\eqref{erpot1}.
Specifically, $f$ and $A_t$ are identified as the real parts of $\mathcal{E}+|\Phi|^2$ and $\Phi$, respectively. 
Furthermore, the components of Eqs.~\eqref{erpot2} and \eqref{erpot1} are
\begin{align}
    \left\{
    \begin{aligned}
        \partial_\rho\omega &= -\rho\frac{\partial_z\left(\Im\mathcal{E}\right)+2\Im\!\left(\Phi^*\partial_z\Phi\right)}{\left(\Re\mathcal{E}+\abs{\Phi}^2\right)^2}, \\
        \partial_z\omega &= +\rho\frac{\partial_\rho\left(\Im\mathcal{E}\right)+2\Im\!\left(\Phi^*\partial_\rho\Phi\right)}{\left(\Re\mathcal{E}+\abs{\Phi}^2\right)^2},
    \end{aligned}
    \right. \label{eq omega}
\end{align}
and
\begin{align}
    \left\{
    \begin{aligned}
        \partial_\rho A_\phi &= - \omega\partial_\rho\left(\Re\Phi\right)+\frac{\rho}{\Re\mathcal{E}+\abs{\Phi}^2}\partial_z\left(\Im\Phi\right), \\
        \partial_z A_\phi &= - \omega\partial_z\left(\Re\Phi\right)-\frac{\rho}{\Re\mathcal{E}+\abs{\Phi}^2}\partial_\rho\left(\Im\Phi\right),
    \end{aligned}
    \right. \label{eq A phi}
\end{align}
The integrability conditions for these equations are guaranteed by the Ernst equations~\eqref{ernsteq}. 
Consequently, integrating Eqs.~\eqref{eq omega} and \eqref{eq A phi} yields the explicit forms of $\omega$ and $A_\phi$.

While the $(a,i)$-components of the Einstein equations \eqref{Rai =} are trivially satisfied under our ansatz, 
the remaining components \eqref{einstein1} reduce to differential equations for $\gamma$. 
These can be recast in the following form:
\begin{align}
    \left\{
    \begin{aligned}
        \partial_\rho\gamma &= \rho\left[\frac{1}{4}\frac{\abs{\partial_\rho\mathcal{E}+2\Phi^*\partial_\rho\Phi}^2-\abs{\partial_z\mathcal{E}+2\Phi^*\partial_z\Phi}^2}{\left(\Re\mathcal{E} + \abs{\Phi}^2\right)^2} - \frac{\abs{\partial_\rho\Phi}^2-\abs{\partial_z\Phi}^2}{\Re\mathcal{E}+\abs{\Phi}^2}
        + \frac{1}{2} (S_{\rho\rho}^{(\mathrm{m})} - S_{zz}^{(\mathrm{m})} )
        \right], \\
        \partial_z\gamma &= 2\rho\left[\frac{1}{4}\frac{\Re\!\left[\left(\partial_\rho\mathcal{E}+2\Phi^*\partial_\rho\Phi\right)\left(\partial_z\mathcal{E}^*+2\Phi\partial_z\Phi^*\right)\right]}{\left(\Re\mathcal{E} + \abs{\Phi}^2\right)^2} - \frac{\Re\!\left[\left(\partial_\rho\Phi\right)\left(\partial_z\Phi^*\right)\right]}{\Re\mathcal{E}+\abs{\Phi}^2}
        +
        \frac{1}{2} S_{\rho z}^{(\mathrm{m})}
        \right].
    \end{aligned}
    \right. \label{gamma}
\end{align}
The integrability condition for $\gamma$ is guaranteed by the Ernst equation~\eqref{ernsteq} and the conservation law of the matter field.

The discussion above can be applied to any matter field that satisfies the condition \eqref{matter1}.
Hereinafter, we mainly consider a massless scalar field $\Psi$, whose Lagrangian is given by
\begin{align}
    \mathcal{L}^{(\mathrm{m})}\left[\Psi\right] &= - \epsilon \frac{1}{\kappa} g^{\mu\nu}\left(\partial_\mu\Psi\right)\left(\partial_\nu\Psi\right).
\end{align}
Here, we introduce a parameter $\epsilon$ which takes $+1$ or $-1$. 
This is because the wormhole constructions discussed in this paper require a phantom field, 
corresponding to $\epsilon=-1$
while the formalism presented in this section is equally applicable to a canonical scalar field corresponding to $\epsilon=+1$. 
The energy momentum tensor of this scalar field is evaluated as
\begin{align}
S_{\mu\nu}^{(\mathrm{m})} &= 2\epsilon\left(\partial_\mu\Psi\right)\left(\partial_\nu\Psi\right).
\end{align}
We assume stationarity and axisymmetry of the scalar field, which implies
\begin{align}
\Psi = \Psi(\rho,z).
\end{align}
Under this assumption, one can see that the energy momentum tensor of the scalar field satisfies our requirement \eqref{matter1}.

The equation of motion for the scalar field is the Klein-Gordon equation, $\nabla^\mu\nabla_\mu\Psi=0$.
Under the symmetries we assumed, this reduces to the Laplace equation on $\mathbb{R}^3$ with axisymmetry,
\begin{align}
D^2 \Psi
= 0. \label{kg1}
\end{align}

To summarize, the dynamical variables of this system are two complex potentials $\mathcal{E}, \Phi$ and the real scalar field $\Psi$, which follow the Ernst and Laplace equations, Eqs \eqref{ernsteq} and \eqref{kg1}. 
Once a solution to the Eqs.~\eqref{ernsteq} and \eqref{kg1} is obtained, all other functions appearing in the metric and gauge potential can be determined. Specifically, $f$ and $A_t$ are obtained directly from the definitions of the potentials, Eqs.~\eqref{erpot3} and \eqref{Phipot5}:
 \begin{align}
 f &= \Re \mathcal{E} + |\Phi|^2, \\
 A_{t} &= \Re \Phi, 
 \end{align}
while $\omega, A_{\phi}$, and $\gamma$ are found by integrating the differential equations \eqref{eq omega}, \eqref{eq A phi}, and \eqref{gamma}.

\subsection{The Harrison transformation} \label{harrison}
The Ernst equations \eqref{ernsteq} can be derived as the Euler-Lagrange equations from the following Lagrangian, with $\mathcal{E}$ and $\Phi$ treated as dynamical variables,
\begin{align}
    L[\mathcal{E},\Phi] &= \rho \left( \frac{\abs{D \mathcal{E}+2\Phi^*D\Phi}^2}{\left(\Re\mathcal{E}+\abs{\Phi}^2\right)^2} - \frac{4\abs{D\Phi}^2}{\Re\mathcal{E}+\abs{\Phi}^2} \right). \label{lagrangian1}
\end{align}
This Lagrangian is invariant under $SU(2, 1)$ transformations.
Here we consider an element of that family~\cite{Harrison1968, Vigano:2022hrg},
\begin{align}
    \left\{
    \begin{aligned}
        \mathcal{E} &\to \frac{c^2\mathcal{E}}{1-2q^*\Phi-\abs{q}^2\mathcal{E}}, \\
        \Phi & \to \frac{c\left(q\mathcal{E}+\Phi\right)}{1-2q^*\Phi-\abs{q}^2\mathcal{E}},
    \end{aligned}
    \right. 
\end{align}
where $c$ and $q$ are real and complex constants, respectively, and the asterisk~${}^*$ denotes the complex conjugate.
Specifically, if a pair of potentials $(\bar{\mathcal{E}}, \bar{\Phi})$ represents a solution to the Ernst equations, then the transformed potentials
\begin{align}
    \left\{
    \begin{aligned}
        \mathcal{E} &= \frac{c^2 \bar{\mathcal{E}}}{1-2q^*\bar{\Phi}-\abs{q}^2\bar{\mathcal{E}}}, \\
        \Phi & =  \frac{c\left(q\bar{\mathcal{E}}+\bar{\Phi}\right)}{1-2q^*\bar{\Phi}-\abs{q}^2\bar{\mathcal{E}}},
    \end{aligned}
    \right. \label{Harrison trns}
\end{align}
are also solutions. 
This is nothing but the solution generating method via a transformation from a seed $(\bar{\mathcal{E}}, \bar{\Phi})$.
This transformation~\eqref{Harrison trns} is referred to as \textit{the Harrison transformation} if the parameter $c$ is set to unity\footnote{
We keep the parameter $c$ arbitrary as we will perform a coordinate rescaling later.
}.
The real parameter $c$ rescales a coordinate,
while the parameter $q$, when starting from a neutral seed, determines the amplitude of the additional electromagnetic field.
More precisely, the real part of $q$ controls the amplitude of the electric field, while the imaginary part governs that of the magnetic field.

The Klein-Gordon equation~\eqref{kg1} does not depend on $\mathcal{E}$ and $\Phi$. 
Therefore, we can reuse the same function as the solution in the transformed system, that is, 
\begin{align}
    \Psi = \bar\Psi.
\end{align}

Substituting $(\mathcal{E}, \Phi, \Psi)$ into Eq.~\eqref{gamma}, one finds 
that the right hand sides are unchanged and, therefore, we obtain the solution $\gamma = \bar{\gamma}$ up to a constant shift.

\subsection{An example: Reissner-Nordstr\"om solution from Schwarzschild seed} \label{schwarzschild}
In this section, we review the generation of the Reissner-Nordstr\"om solution from the Schwarzschild seed via the Harrison transformation.
This procedure serves to illustrate how a charged multi-sheet wormhole solution can be generated from the neutral seed.
The Schwarzschild solution is written in the spherical coordinates $(t, r, \theta, \phi)$ as
\begin{align}
    \d s^2 &= -f_\mathrm{Sch}(r)\d t^2 +\frac{1}{f_\mathrm{Sch}(r)}\d r^2 +r^2\d\Omega^2,
\end{align}
where $f_\mathrm{Sch}(r) \coloneqq 1-2m/r$ and $\d\Omega^2\coloneqq\d\theta^2+\sin^2\theta\d\phi^2$ is the metric of the unit two-sphere.
In a static region, namely $r>2m$, we can introduce the Weyl coordinates $(\rho, z)$ by
\begin{align}
    \left\{
    \begin{aligned}
        \rho &= \sqrt{r(r-2m)}\sin\theta, \\
        z &= (r-m)\cos\theta.
    \end{aligned}
    \right. \label{schwarzschild1}
\end{align}
The line element is then rewritten in the Weyl-Papapetrou form,
\begin{align}
    \d s^2 &= -f_\mathrm{Sch}(\rho, z)\d t^2 + \frac{1}{f_\mathrm{Sch}(\rho, z)}\left[\e^{2\gamma_\mathrm{Sch}(\rho, z)}\left(\d\rho^2+\d z^2\right)+\rho^2\d\phi^2\right],
\end{align}
where the functions with the subscript ``Sch'', denoting the quantities in Eq.~\eqref{WP} for the Schwarzschild solution, are given by
\begin{align}
    f_{\mathrm{Sch}} &= \frac{\mu_{-m}}{\mu_{+m}}, \qquad \omega_{\mathrm{Sch}} = 0, \qquad \e^{2\gamma_{\mathrm{Sch}}} = \frac{\left(\rho^2+\mu_{+m}\mu_{-m}\right)^2}{\left(\rho^2+\mu_{+m}^2\right)\left(\rho^2+\mu_{-m}^2\right)}, \label{Sch in WP}
\end{align}
and the functions $\mu_{\pm m}$ are defined by
\begin{align}
    \mu_{\pm m} &\coloneqq -(z\mp m)+\sqrt{\rho^2+(z\mp m)^2}. \label{soliton1}
\end{align}
We call them soliton functions.
Note that $\ln\mu_{\pm m}$ are harmonic functions satisfying $D^2 \ln\mu_{\pm m} =0$.
As used as the definitions of $\mu_{\pm m}$ in Refs.~\cite{Belinski1978, Belinski1979, Belinski:2001ph}, $\mu_{\pm m}$ satisfy
\begin{align}
    \partial_\rho\mu_{\pm m} &= \frac{2\rho\mu_{\pm m}}{\rho^2+\mu^2_{\pm m}}, \qquad \partial_z \mu_{\pm m} = -\frac{2\mu^2_{\pm m}}{\rho^2+\mu^2_{\pm m}}. \label{soliton3}
\end{align}
From the expression of the Schwarzschild solution in the Weyl-Papapetrou form~\eqref{Sch in WP}, we construct the Ernst potentials.
The corresponding potentials for the Schwarzschild seed become
\begin{align}
    \mathcal{E}_\mathrm{Sch} = f_{\mathrm{Sch}} = \frac{\mu_{-m}}{\mu_{+m}}, \qquad \Phi_\mathrm{Sch} = 0. \label{schpot}
\end{align}

The Reissner-Nordstr\"{o}m solution can be generated from the Schwarzschild seed
by applying the Harrison transformation \eqref{Harrison trns}.
The resulting Ernst and electromagnetic potentials for the Reissner-Nordstr\"{o}m solution are 
\begin{align}
    \mathcal{E}_{\mathrm{RN}} = \frac{c^2\mathcal{E}_\mathrm{Sch}}{1-\abs{q}^2\mathcal{E}_\mathrm{Sch}} , \qquad \Phi_{\mathrm{RN}} = \frac{cq\mathcal{E}_\mathrm{Sch}}{1-\abs{q}^2\mathcal{E}_\mathrm{Sch}}, \label{bhpot1}
\end{align}
where $c$ is a real parameter while $q$ is a complex parameter satisfying $\abs{q}<1$.
As a preliminary note, we shall later set $c=1-\abs{q}^2$ so that the coordinate time $t$ corresponds to the normalized time at spatial infinity.
We can reconstruct the metric functions and the gauge field from these potentials as
\begin{align}
    f_{\mathrm{RN}} &= \Re\mathcal{E}_{\mathrm{RN}} + \abs{\Phi_{\mathrm{RN}}}^2 = \frac{c^2f_{\mathrm{Sch}}}{\left(1-\abs{q}^2f_{\mathrm{Sch}}\right)^2}, \\
    \tilde\omega_{\mathrm{RN}} &= \Im\mathcal{E}_{\mathrm{RN}} = 0, \\
    (A_{\mathrm{RN}})_t &= \Re\Phi_{\mathrm{RN}} = \left(\Re q\right)\frac{cf_{\mathrm{Sch}}}{1-|q|^2f_{\mathrm{Sch}}}, \\
    (\tilde A_{\mathrm{RN}})_\phi &= \Im\Phi_{\mathrm{RN}} = \left(\Im q\right)\frac{cf_{\mathrm{Sch}}}{1-|q|^2f_{\mathrm{Sch}}}.
\end{align}
Since the right-hand sides of Eq.~\eqref{eq omega} vanish, it follows immediately that $\omega_{\mathrm{RN}}=0$. 
Meanwhile, $(A_{\mathrm{RN}})_{\phi}$ can be determined by solving  
Eq.~\eqref{eq A phi}, which is expressed as
\begin{align}
    \left\{
    \begin{aligned}
            \partial_\rho (A_{\mathrm{RN}})_\phi &= +\frac{\left(\Im q\right)}{c}\rho\partial_z\ln f_\mathrm{Sch}, \\
    \partial_z (A_{\mathrm{RN}})_\phi &= -\frac{\left(\Im q\right)}{c}\rho\partial_\rho\ln f_\mathrm{Sch}.
    \end{aligned}
    \right.
\end{align}
Recalling that the solitons $\mu_\pm$ in the potential $f_\mathrm{Sch}$ satisfy the relations~\eqref{soliton3}, we find
\begin{align}
    (A_{\mathrm{RN}})_\phi &= -\frac{\left(\Im q\right)}{c}\left(\mu_{-m}-\mu_{+m}\right).
\end{align}

The Harrison transformation does not modify the function $\gamma$, resulting in $\gamma_{\mathrm{RN}} = \gamma_{\mathrm{Sch}}$.
Consequently, the expression for the generated solution reads
\begin{align}
    \d s^2_{\mathrm{RN}} &= - \frac{c^2f_{\mathrm{Sch}}}{\left(1-\abs{q}^2f_{\mathrm{Sch}}\right)^2}\d t^2 +  \frac{\left(1-\abs{q}^2f_{\mathrm{Sch}}\right)^2}{c^2f_{\mathrm{Sch}}}\left[\frac{\left(\rho^2+\mu_{+m}\mu_{-m}\right)^2}{\left(\rho^2+\mu_{+m}^2\right)\left(\rho^2+\mu_{-m}^2\right)}\left(\d\rho^2+\d z^2\right) + \rho^2\d\phi^2\right], \\
    (A_{\mathrm{RN}})_\mu\d x^\mu &= \left(\Re q\right)\frac{cf_{\mathrm{Sch}}}{1-|q|^2f_{\mathrm{Sch}}}\d t - \frac{\left(\Im q\right)}{c}\left(\mu_{-m}-\mu_{+m}\right)\d\phi.
\end{align}
By choosing $c=1-\abs{q}^2$ so that the norm of the timelike Killing vector $\partial_t$ becomes unity at spatial infinity, the final form of the Reissner-Nordström solution in the Weyl-Papapetrou form is given by
\begin{align}
    \d s_\mathrm{RN}^2 &= - \frac{\left(1-|q|^2\right)^2f_{\mathrm{Sch}}}{\left(1-\abs{q}^2f_{\mathrm{Sch}}\right)^2}\d t^2 +  \frac{\left(1-\abs{q}^2f_{\mathrm{Sch}}\right)^2}{\left(1-|q|^2\right)^2f_{\mathrm{Sch}}}\left[\frac{\left(\rho^2+\mu_{+m}\mu_{-m}\right)^2}{\left(\rho^2+\mu_{+m}^2\right)\left(\rho^2+\mu_{-m}^2\right)}\left(\d\rho^2+\d z^2\right) + \rho^2\d\phi^2\right], \\
    (A_{\mathrm{RN}})_\mu\d x^\mu &= \left(\Re q\right)\frac{\left(1-|q|^2\right)f_{\mathrm{Sch}}}{1-|q|^2f_{\mathrm{Sch}}}\d t - \frac{\left(\Im q\right)}{1-\abs{q}^2}\left(\mu_{-m}-\mu_{+m}\right)\d\phi.
\end{align}

Let us confirm that this solution is identical to the standard form of the Reissner-Nordstr\"{o}m solution.
By transforming back to the original coordinates $(r, \theta)$ via Eq.~\eqref{schwarzschild1} and defining a rescaled radial coordinate $R$ as
\begin{align}
    R &\coloneqq \frac{1-\abs{q}^2f_{\mathrm{Sch}}}{1-|q|^2}r = r + 2m\frac{\abs{q}^2}{1-|q|^2},
\end{align}
we obtain the solution in the conventional expressions,
\begin{align}
    \d s^2 &= -\left(1-\frac{2M}{R} + \frac{Q_\mathrm{e}^2+Q_\mathrm{m}^2}{R^2}\right)\d t^2 + \left(1-\frac{2M}{R} + \frac{Q_\mathrm{e}^2+Q_\mathrm{m}^2}{R^2}\right)^{-1}\d R^2 + R^2\d\Omega^2 \\
    A_\mu\d x^\mu &= - \frac{Q_\mathrm{e}}{R}\d t - Q_\mathrm{m}\cos\theta\d\phi,
\end{align}
up to $U(1)$ gauge transformation. 
For a more transparent physical interpretation, it is instructive to express the electromagnetic field in terms of the gauge-invariant field strength tensor rather than the gauge potential,
\begin{align}
    \frac{1}{2}F_{\mu\nu}\d x^\mu\wedge\d x^\nu &= -\frac{Q_\mathrm{e}}{R^2}\d t\wedge\d R + Q_\mathrm{m}\sin\theta\d\theta\wedge\d\phi.
\end{align}
The gravitational and electromagnetic charges of the system, as measured at spatial infinity, are identified as follows,
\begin{alignat}{3}
    \text{ADM mass}&\qquad&M &= \frac{1+\abs{q}^2}{1-|q|^2}m, \\
    \text{electric charge}&\qquad&Q_\mathrm{e} &= \frac{2\left(\Re q\right)}{1-|q|^2}m, \\
    \text{magnetic charge}&\qquad&Q_\mathrm{m} &= \frac{2\left(\Im q\right)}{1-|q|^2}m.
\end{alignat}
Note that these new parameters satisfy 
\begin{align}
    M^2 - \left(Q_\mathrm{e}^2+Q_\mathrm{m}^2\right) &= m^2 > 0,\label{mass0}
\end{align}
indicating that the generated Reissner-Nordstr\"om solutions are always sub-extremal.

\section{
Multi-sheet wormhole solutions
} \label{charged}
In this section, we apply the Harrison transformation to the multi-sheet wormholes presented in our previous paper~\cite{Makita:2025bao}.
We see some analogies with the black hole cases in the process of generating these wormhole solutions.
To support the wormhole structure, we consider a massless \textit{phantom} scalar field as a matter source, corresponding to $\epsilon=-1$.

\subsection{
Neutral multi-sheet wormhole solutions
} \label{prelim}
To generate charged multi-sheet wormhole solutions, we consider the neutral multi-sheet wormhole solutions found in our previous paper~\cite{Makita:2025bao} as seeds.
Our multi-sheet solutions connect an even number, say $2n$, of spatial infinities,
and are described by the corresponding number of $\rho z$ sheets connected through the branch cuts, $0<\rho<\alpha, z=0$.
These sheets are categorized into two types, labeled by the sign $(\pm)$. 
In each sheet, the metric and the scalar field can be expressed as
\begin{align}
    \d \bar{s}_\pm^2 &= -\bar{f}_\pm(\rho, z)\d t^2 + \frac{1}{\bar{f}_\pm(\rho, z)}\left\{A\left[1-\frac{\rho^2}{4\alpha^2}(1-w(\rho, z))^2\right]^{-B}\left(\d\rho^2+\d z^2\right)+\rho^2\d\phi^2\right\}, \label{g NWH}
\end{align}
and
\begin{align}
    \bar{\Psi}_\pm &= \frac{\Q}{2\alpha}\ln\mu_\pm(\rho, z), \label{Psi NWH}
\end{align}
where
\begin{align}
    \bar{f}_\pm &\coloneqq \left(\mu_\pm\right)^{-\frac{m}{\alpha}}, \label{deff1} \\
    \ln\mu_+ &\coloneqq \Arccos{w}, \label{defmu1} \\
    \ln\mu_- &\coloneqq 2\pi-\Arccos{w}, \label{defmu2} \\
    w &\coloneqq -\frac{z^2+\alpha^2}{\rho^2} + \frac{\sqrt{\left(\rho^2+z^2-\alpha^2\right)^2+4\alpha^2z^2}}{\rho^2}, \label{defw} \\
    B &\coloneqq \frac{\Q^2-m^2}{\alpha^2}. \label{defB1} 
\end{align}
Here, $m, \Q, \alpha$ and $A$ are free parameters. 
Specifically, $m$ is the ADM mass, $\alpha$ is the throat radius, and $\Q$ is the scalar charge observed at spatial infinity in \ps sheet.
The remaining parameter $A$ is fixed later by the regularity conditions.
For simplicity, we assume that these parameters are all positive.
The metric \eqref{g NWH} and the scalar field \eqref{Psi NWH} are local solutions to the Einstein and the Klein-Gordon equations, provided that the phantom field is chosen, that is, $\epsilon = -1$.
We note that the function $w$ in Eq.~\eqref{defw} approaches to unity at each asymptotic region while $-1$ at the throat of the wormhole.
Then $\ln\mu_+$ corresponding to \ps sheet takes the value between 0 and $\pi$, while $\ln\mu_-$ corresponding to \ms sheet takes values between $\pi$ and $2\pi$.

From the expression of the metric, the Ernst potential reads
\begin{align}
    \bar{\mathcal{E}} = \bar{f}_{\pm} = \left(
    \mu_{\pm}
    \right)^{- \frac{m}{\alpha}}. \label{whpot}
\end{align}
Since this seed solution lacks an electromagnetic field, we simply have
\begin{align}
     \bar{\Phi} = 0.
\end{align}

As shown in Ref.~\cite{Makita:2025bao}, the solitons $\mu_{\pm}$ can be expressed in a similar way to the Schwarzschild case as follows:
\begin{align}
    \mu_+ &= \left(\frac{\mu_{-\I\alpha}}{\mu_{+\I\alpha}}\right)^\I, \qquad \mu_- = \e^{2\pi}\left(\frac{\mu_{+\I\alpha}}{\mu_{-\I\alpha}}\right)^\I, \label{soliton4}
\end{align}
where $\mu_{\pm\I\alpha}$ are solitons obtained by replacing $m$ in Eq.~\eqref{soliton1} with $\I \alpha$, that is,
\begin{align}
\mu_{\pm\I\alpha} &\coloneqq -(z\mp\I\alpha) + \sqrt{\rho^2+(z\mp\I\alpha)^2}. \label{soliton2}
\end{align}
The powers of complex functions are defined using the principal branch of the logarithm. In this paper, we take the branch cut along the negative real axis and choose the argument to lie in $[-\pi,\pi)$.

If one imposes the global regularity, the parameter space is constrained.
Based on Ref.~\cite{Makita:2025bao},
the requirement for the absence of
conical singularities on the axis as well as the asymptotic flatness independently result in $A=1$.
Furthermore, the regularity across the sheets requires that the value of $B$ is determined by the number of sheets,
\begin{align}
    B &= \frac{\Q^2 - m^2}{\alpha^2} = \frac{2n-1}{n} \label{constraint1},
\end{align}
where $n$ is the number of pairs of \ps and \ms sheets. Namely, this solution connects individual $2n$ asymptotic regions. 
This constraint on $B$ establishes a relation among $P$, $m$ and $\alpha$, 
implying that one of these parameters, namely, the scalar charge, the ADM mass and the throat radius, is no longer a free parameter for a regular wormhole. 
However, we do not impose these constraints on $A$ and $B$ at the level of the seed solution. 
The reason is as follows:
While the Harrison transformation leaves the differential equations invariant, which ensures that the generated solution remains a local solution to the field equations, it does not necessarily preserve the boundary conditions. 
Specifically, the regularity on the boundaries of the $(\rho, z)$-space is not guaranteed to be invariant under the transformation. 
In past studies of black object construction~\cite{Mishima:2005id,Tomizawa:2005wv,Tomizawa:2006vp,Yazadjiev:2006hw,Elvang:2007rd,Evslin:2007fv,Izumi:2007qx,Elvang:2007hs,Iguchi:2011qi,Rocha:2011vv,Tomizawa:2022qyd,Suzuki:2023ine,Suzuki:2024coe,Tomizawa:2024tkh,Suzuki:2024phv}, 
it has been observed that seeds possessing singularities on the boundaries can sometimes generate globally regular solutions, and vice versa. 
Therefore, we postpone the imposition of the regularity conditions until after the transformation, 
and subsequently verify the global regularity of the resulting charged multi-sheet wormhole.
Hence, we leave all four parameters, $m, \alpha, \Q$, and $A$, as free parameters for the time being, and derive the constraints required for the global regularity of the generated solution after performing the transformation.

\subsection{Charged multi-sheet wormhole solutions}

Let us generate charged multi-sheet wormhole solutions from the neutral one explained above.
In what follows, we omit the plus-minus subscript whenever it is clear from context.
The Harrison transformation \eqref{Harrison trns} can be performed as
\begin{align}
    \mathcal{E} = \frac{c^2\bar{\mathcal{E}}}{1-\abs{q}^2\bar{\mathcal{E}}}, \qquad \Phi = \frac{cq\bar{\mathcal{E}}}{1-\abs{q}^2\bar{\mathcal{E}}}, \label{whpot1}
\end{align}
where $c$ is a real constant while $q$ is an arbitrary complex parameter satisfying $\abs{q}<1$.
As explained before, the functions $f$ and $A_t$ are able to be directly obtained from these potentials as 
\begin{align}
    f &= \Re \mathcal{E} + \abs{\Phi}^2 = \frac{c^2 \bar{f}}{\left(1-\abs{q}^2 \bar{f}\right)^2}, \label{deff2} \\
    A_{t} &= \Re \Phi = \left(\Re q\right)\frac{ c\bar{f}}{1-\abs{q}^2\bar{f}},
\end{align}
where $\bar{f}$ is the seed potential given in Eqs.~\eqref{deff1}.
The remaining components, $\omega$ and $A_{\phi}$, can be obtained by integrating the differential equations.
It is obvious that $\mathcal{E}$ and $\Phi^{*} \partial_{i}\Phi$
are real, so we find $\omega=0$, \textit{i.e.}, the system is static.
The magnetic potential $A_\phi$ can be reconstructed from the differential equations,
\begin{align}
    \left\{
    \begin{aligned}
        \partial_\rho A_\phi &= +\frac{\left(\Im q\right)}{c}\rho\partial_z\ln\bar{\mathcal{E}}, \\
        \partial_z A_\phi &= -\frac{\left(\Im q\right)}{c}\rho\partial_\rho\ln\bar{\mathcal{E}}.
    \end{aligned}
    \right. \label{whpot2}
\end{align}
The solution is
\begin{align}
    A^\pm_\phi &= -\frac{\left(\Im q\right)}{c}\left(-\frac{m}{\alpha}\I\right)\left(\mu_{\mp\I\alpha}-\mu_{\pm\I\alpha}\right).
\end{align}
As explained in the previous section, the Harrison transformation leaves $\gamma$ unchanged. 
Thus, we obtain
\begin{align}
\e^{2\gamma} = \e^{2\bar{\gamma}} = A\left[1-\frac{\rho^2}{4\alpha^2}\left(1-w(\rho, z)\right)^2\right]^{-B}.
\end{align}

In the following, we fix the parameter $c$ as $c=1-\abs{q}^2$. Under this condition, our solution can be summarized as follows:
\begin{align}
    \d s_\pm^2 &= -f_\pm(\rho, z)\d t^2 + \frac{1}{f_\pm(\rho, z)}\left\{A\left[1-\frac{\rho^2}{4\alpha^2}\left(1-w(\rho, z)\right)^2\right]^{-B}\left(\d\rho^2+\d z^2\right)+\rho^2\d\phi^2\right\}, \label{charged1a} \\
    A_\mu^\pm(\rho, z)\d x^\mu &= \left(\Re q\right)\frac{\left(1-\abs{q}^2\right)\bar{f}_\pm(\rho, z)}{1-\abs{q}^2\bar{f}_\pm(\rho, z)}\d t - \frac{\left(\Im q\right)}{1-\abs{q}^2}\left(-\frac{m}{\alpha}\I\right)\left(\mu_{\mp\I\alpha}-\mu_{\pm\I\alpha}\right)\d\phi, \label{charged1b} \\
    \Psi_\pm(\rho, z) &= \frac{\Q}{2\alpha}\ln\mu_\pm. \label{charged1c}
\end{align}
Here, we assume that the parameters $A$ and $B$ are real, while $q$ is complex satisfying $\abs{q}<1$.
The global regularity conditions, which fix the values of $A$ and $B$, will be discussed in detail in Sec.~\ref{regularities}.

The physical quantities evaluated at spatial infinity on \ps sheet are identified as follows:
\begin{alignat}{3}
    \text{ADM mass}&\qquad&M &= \frac{1+\abs{q}^2}{1-\abs{q}^2}m, \label{params1} \\
    \text{electric charge}&\qquad&Q_\mathrm{e} &= \frac{2\left(\Re q\right)}{1-\abs{q}^2}m, \\
    \text{magnetic charge}&\qquad&Q_\mathrm{m} &= \frac{2\left(\Im q\right)}{1-\abs{q}^2}m, \label{params2} \\
    \text{scalar charge}&\qquad&\Q, \\
    \text{throat radius}&\qquad&\alpha,
\end{alignat}
where  $m$ in the right hand sides is the ADM mass of the neutral seed solution.
Details of evaluation are given in Sec.~\ref{asymptotic}.
The solution is characterized by five physical parameters: $M, Q_\mathrm{e}, Q_\mathrm{m}, \Q$, and $\alpha$, which are controlled by five model parameters, $m$, $\Re q$, $\Im q$, $P$ and $\alpha$. 

Eliminating $q$ from Eqs.~\eqref{params1}--\eqref{params2}, one finds
\begin{align}
    M^2 - Q^2 &= m^2 > 0, \label{mass1}
\end{align}
where we introduced the electromagnetic charge parameter $Q \! \coloneqq \! \sqrt{Q_\mathrm{e}^2+Q_\mathrm{m}^2}$ for simplicity.
The relation \eqref{mass1}
implies that the system satisfies a condition analogous to the sub-extremal condition for black holes given in Eq.~\eqref{mass0}. 
In addition, using Eqs.~\eqref{params1}--\nobreak\eqref{params2} and the assumptions $m>0$ and $\abs{q}<1$, one can derive a relation,
\begin{align}
    \abs{q} &= \frac{M-\sqrt{M^2-Q^2}}{Q}. \label{eqforq}
\end{align}
These relations allow for the solution to be expressed entirely in terms of physical quantities, 
eliminating the parameter $m$ and $|q|$.

Let us rewrite the solution in terms of the physical quantities.
First, by substituting Eq.~\eqref{mass1} into Eq.~\eqref{defB1}, the index $B$ in the metric \eqref{charged1a} can be expressed in terms of these physical parameters as
\begin{align}
    B &= \frac{\Q^2+Q^2-M^2}{\alpha^2}. \label{constraint2}
\end{align}

Next, let us express $f_{\pm}$ by the physical parameters.
Let us introduce a new variable $\Theta$ containing $\bar f$ in Eq.~\eqref{deff1} by
\begin{align}
    \Theta_\pm &\coloneqq -\frac{1}{2}\ln\left[\frac{(M-\sqrt{M^2-Q^2})^2}{Q^2}\bar{f}_\pm\right] 
    = \frac{\sqrt{M^2-Q^2}}{2\alpha}\ln\mu_\pm + \ln\left(\frac{M+\sqrt{M^2-Q^2}}{Q}\right) .\label{deftheta1}
\end{align}
Then, eliminating $|q|$ by Eq.~\eqref{eqforq}, we can express the metric function $f$ as
\begin{align}
    f &= \frac{\left(1-|q|^2\right)^2 \bar{f}}{\left(1-|q|^2 \bar{f}\right)^2} = \frac{M^2-Q^2}{Q^2}\frac{1}{\sinh^2\Theta}. \label{deff3}
\end{align}

Furthermore, by eliminating $q$ from Eq.~\eqref{charged1b}, we find
\begin{align}
    A_t &= Q_\mathrm{e}\frac{\sqrt{M^2-Q^2}}{Q^2}\frac{\e^{-\Theta}}{\sinh\Theta}, \\
    A_\phi^\pm &= \frac{Q_\mathrm{m}}{2\alpha}\I\left(\mu_{\mp\I\alpha}-\mu_{\pm\I\alpha}\right). \label{magnetic1}
\end{align}
These expressions enable us to write down the electromagnetic fields in a gauge-invariant form, \textit{i.e.}, the field strength tensor $F_{\mu\nu}$. 
The derivatives of the potentials are calculated as
\begin{align}
    \partial_iA_t &= -\frac{Q_\mathrm{e}}{2\alpha}  f \left(\partial_i\ln\mu\right), \label{faradayt} \\
    \partial_iA_\phi^\pm &= +\frac{Q_\mathrm{m}}{2\alpha}\I\left(\partial_i\mu_{\mp\I\alpha}-\partial_i\mu_{\pm\I\alpha}\right). \label{faradayphi}
\end{align}
It follows that
\begin{align}
    \frac{1}{2}F_{\mu\nu}^\pm\d x^\mu\wedge\d x^\nu &= \frac{Q_\mathrm{e}}{2\alpha}f_\pm\left(\partial_i\ln\mu_\pm\right)\d t\wedge\d y^i + \frac{Q_\mathrm{m}}{2\alpha}\I\left(\partial_i\mu_{\mp\I\alpha}-\partial_i\mu_{\pm\I\alpha}\right)\d y^i\wedge\d\phi, \label{faraday}
\end{align}
where $y^i$ denotes the spatial coordinates $(\rho,z)$.

Let us comment on the neutral limit of our solution.
The expression for $f$, Eq.~\eqref{deff3}, appears to be singular in the neutral limit $Q \to 0$. 
However, the product $Q \sinh\Theta$ remains finite in the limit $Q \to 0$, and thus, 
this limit correctly recovers the neutral seed solution given in Eq.~\eqref{deff1}.
Therefore, we obtain 
\begin{align}
    \lim_{Q\to0} f &= \mu^{-\frac{M}{\alpha}} = \bar{f}.
\end{align}

The condition~\eqref{mass1} provides a ``sub-extremal'' condition, $M^2>Q^2$.
We extend the solutions beyond this limit in Sec.~\ref{extension}.

\subsection{Regularity analyses} \label{regularities}
In this subsection, we analyze global regularities of the charged multi-sheet wormholes generated in the previous subsection, following the procedure established in Ref.~\cite{Makita:2025bao}. 
Since the Harrison transformation leaves the form of the field equations invariant, 
the generated solution automatically satisfies them locally. 
However, the regularity at the boundary of the $\rho z$ plane is not automatic and must be checked separately.
Specifically, we must verify the regularities in three regions, the regularity across the cut, at the edge, and on the axis.
As our solutions are composed of the solitons $\mu_\pm$ introduced in Ref.~\cite{Makita:2025bao}, 
the verification follows a process largely identical to that in the neutral case.
Below, we provide an outline of these analyses, with the technical details summarized in Appendix~\ref{analyses} (see also Ref.~\cite{Makita:2025bao}).

\subsubsection{Regularity across the cut} \label{cut}
As shown in Ref.~\cite{Makita:2025bao}, the solitons $\mu_\pm$ on each sheet have a ``cusp'' segment ($0<\rho<\alpha, \, z=0$) in the $\rho z$ plane. 
The wormhole solution is then constructed by gluing different types of sheets along this cut.
Therefore, it is necessary to verify the regularity across the glued region.
In Ref.~\cite{Makita:2025bao} it is established that the solitons $\mu_\pm$ are smoothly connected.
For the charged solution given in Eqs.~\eqref{charged1a}-\eqref{charged1c}, 
this regular structure remains unchanged since the functions are composed of the same solitons $\mu_\pm$.

\subsubsection{Regularity at the edge of the cut} \label{edge}
A naive gluing of multiple sheets generally introduces a conical singularity at the edge of the cut.

Let us first confirm the absence of a conical singularity.
Suppose we have $2n$ sheets.
In order to describe these $2n$ glued sheets as a single patch around the edge, a \textit{connected} polar coordinate on each sheet is introduced in Ref.~\cite{Makita:2025bao} as
\begin{align}
    \left\{\,
    \begin{aligned}
        \rho - \alpha &= -\tilde r^{2n}\cos\,(2n\tilde\theta), \\
        z &= -\tilde r^{2n}\sin\,(2n\tilde\theta).
    \end{aligned}
    \right. \label{polar1}
\end{align}
We assign the angular domain for each sheet as
\begin{align}
    \left\{\,
    \begin{aligned}
        \frac{2(k-1)}{n}\pi &\leq \tilde\theta < \frac{2k-1}{n}\pi \quad & &\text{on sheet} \ (+) {}_k, \\
        \frac{2k-1}{n}\pi &\leq \tilde\theta < \frac{2k}{n}\pi & &\text{on sheet} \ (-) {}_k,
    \end{aligned}
    \right.
\end{align}
where we number the sheets of each type with a subscript $k$ ($k=1, 2, \dots, n$), and the total angle of the glued sheets becomes $2\pi$.
As examined in Ref.~\cite{Makita:2025bao}, the $(i, j)$-block of the metric around the edge can be expressed in the polar coordinate~\eqref{polar1} as
\begin{align}
   \frac{1}{f}\e^{2\gamma}\left(\d\rho^2+\d z^2\right) 
    &\approx (\text{constant}) \times \tilde r^{2(2n-1-nB)}\left(\d\tilde r^2+\tilde r^2\d\tilde\theta^2\right), \label{conical}
\end{align}
where $\approx$ means that the higher order in $\tilde r$ is ignored.
Since the range of $\tilde\theta$ is $2\pi$, the regularity condition around the edge requires the overall factor in Eq.~\eqref{conical} to be independent of $\tilde r$, which implies
\begin{align}
    B &= \frac{2n-1}{n}. \label{sheetnumber1}
\end{align}

Next, we examine the regularity of the functions, $f$, $A_t$, $A_\phi$ and $\Psi$, which ensures that they properly satisfy the field equations at the edge.
We have already verified the regularity of the solitons $\ln\mu_\pm$ in our previous paper~\cite{Makita:2025bao} (see also Sec.~\ref{app: edge}), 
which implies the regularity of the functions, $f, A_t$ and $\Psi$.
Thus, the only remaining function to consider is $A_\phi$, which describes the magnetic potential.
As shown in Eq.~\eqref{A phi in r}, $A_\phi$, given in Eq.~\eqref{magnetic1}, can be expressed in terms of $\tilde r$, which is regular in the limit $\tilde r\to0$.

Referring to Eq.~\eqref{constraint2}, this regularity condition gives a constraint among the physical parameters of the system, $M, Q_\mathrm{e}, Q_\mathrm{m}, \Q$, $\alpha$, and the number of sheets $2n$ as
\begin{align}
    \frac{\Q^2 + Q^2 - M^2}{\alpha^2} &= \frac{2n-1}{n}, \label{regularity1}
\end{align}
noting that $Q^2 = Q_\mathrm{e}^2+Q_\mathrm{m}^2$.
This is the charged generalization of the relation \eqref{constraint1}.
This regularity constraint also implies that the electromagnetic charges $Q_\mathrm{e}$ and $Q_\mathrm{m}$ have the same effects as the phantom scalar charge $P$, which carries the negative energy, from the viewpoint of this regularity condition.

\subsubsection{Regularity on the axis} \label{axis}
The regularity analysis on the axis of symmetry, $\rho=0$, can be performed by taking the limit $\rho\to0$ in the line element,
\begin{align}
    \left.\d s^2_\pm\right|_\text{axis} &= \frac{1}{f_\pm}\left\{A\left[1-\frac{\rho^2}{4\alpha^2}(1-w(\rho, z))^2\right]^{-B}\d\rho^2+\rho^2\d\phi^2\right\} + (\text{regular terms}).
\end{align}
Here, the range of $\phi$ is assumed to be $2\pi$.
Combining with the results in Ref.~\cite{Makita:2025bao} and Sec.~\ref{app: axis}, the line element in the vicinity of the axis is given as
\begin{align}
    \left.\d s_\pm^2\right|_\text{axis} &\approx (\textrm{constant}) \times \left(A\d\rho^2+\rho^2\d\phi^2\right) + (\text{regular terms}).
    \label{axisreg}
\end{align}
Since the prefactor behaves as a constant with respect to $\rho$ and the azimuthal angle $\phi$ runs from 0 to $2\pi$, this result shows that no conical singularities arise on the axis of rotation if and only if $A=1$ holds.

\subsubsection{Asymptotic flatness} \label{asymptotic}
We examine the multi-sheet wormhole with asymptotic flatness.
Details of this analysis are presented in Sec.~\ref{app: asym}.
To verify an asymptotic behavior at spatial infinity, we expand the functions in terms of
\begin{align}
    \epsilon \coloneqq \frac{\alpha}{\sqrt{\rho^2+z^2}},
\end{align}
which approaches zero at spatial infinity.
In the \ps sheet, introducing normalized coordinates,
\begin{align}
    \rho = \rp\sin\theta, \qquad z = \rp\cos\theta, \label{polar2p}
\end{align}
the line element~\eqref{charged1a} reduces to Eq.~\eqref{params3}, which indicates the asymptotic flatness requires $A=1$.
In addition, Eq.~\eqref{params3} shows that the ADM mass observed at this spatial infinity is actually given by $M$ defined by Eq.~\eqref{params1}.
The same applies to the Maxwell field.
Eq.~\eqref{params4} shows that $Q_{\mathrm{e}}$ and $Q_{\mathrm{m}}$ respectively correspond to the electric and magnetic charges observed at this spatial infinity.
These results are completely the same as those of black holes in Sec.~\ref{schwarzschild}.

Similar operations can be performed for the \ms sheet.
We introduce a rescaled time $t_{(-)}$ and spherical coordinates $(\rm, \thm)$ in the sheet by
\begin{align}
    \begin{aligned}
        t &= \left(\frac{Q}{\sqrt{M^2-Q^2}}\sinh\X\right)t_{(-)}, \\
        \rho &= \left(\frac{\sqrt{M^2-Q^2}}{Q}\frac{1}{\sinh\X}\right)\rm\sin\thm, \qquad z = \left(\frac{\sqrt{M^2-Q^2}}{Q}\frac{1}{\sinh\X}\right)\rm\cos\thm,
    \end{aligned}
    \label{polar2m}
\end{align}
where we defined a quantity $\X$ for short notation by
\begin{align}
    \X &\coloneqq \frac{\sqrt{M^2-Q^2}}{\alpha}\pi + \ln\left(\frac{M+\sqrt{M^2-Q^2}}{Q}\right). \label{defX}
\end{align}
Then, we obtain the results similar to those for \ps;
From Eqs.~\eqref{params5} and \eqref{params6}, 
the physical quantities at \ms infinity, namely ADM mass $M_{(-)}$, electric charge $Q_{\mathrm{e}(-)}$, and magnetic charge $Q_{\mathrm{m}(-)}$, are identified with $-Q\cosh\X, -Q_\mathrm{e}$ and $-Q_\mathrm{m}$, respectively.
Indeed, $A=1$ is required for the \ms sheets to ensure asymptotic flatness at this spatial infinity as well.

\section{Extension beyond ``extremal'' limit} \label{extension}

In the previous section, charged multi-sheet wormholes are constructed from the neutral seeds by the Harrison transformation. 
The condition \eqref{mass1}, which is analogous to the sub-extremal condition for black holes, is required in the construction. 
The solution can be extended through analytical continuation, and then, 
the regular solutions exist even in the region $M^2-Q^2\le0$. 
By analogy with black hole cases, we refer to the case with equality as ``\textit{extremal},'' while the case with $M^2 - Q^2 < 0$ is termed ``\textit{overcharged},'' even though these names do not necessarily coincide with the charge bounds of the wormhole itself.
In the following, we first examine the regular overcharged wormholes and then analyze the extremal case.

\subsection{``Overcharged'' case} \label{overcharged}
By simply replacing $\sqrt{M^2-Q^2}$ with $\I\sqrt{Q^2-M^2}$ in Eq.~\eqref{deftheta1}, we have
\begin{align}
    \Theta &= \frac{\sqrt{M^2-Q^2}}{2\alpha}\ln\mu + \ln\left(\frac{M+\sqrt{M^2-Q^2}}{Q}\right) \\
    &= \I\left[\frac{\sqrt{Q^2-M^2}}{2\alpha}\ln\mu + \Arccos\left(\frac{M}{Q}\right)\right] \eqqcolon \I\tilde\Theta, \label{deftheta2}
\end{align}
where we defined an ``angle'' $\tilde\Theta$ for later convenience.
It follows directly from above that
\begin{align}
    f &= \frac{Q^2-M^2}{Q^2}\frac{1}{\sin^2\tilde\Theta},
\end{align}
and therefore $f$ is still real for the overcharged case.

It is obvious that $f$ is singular if there is a point where $\tilde\Theta=k\pi$ 
for an integer $k$.
Equations~\eqref{defmu1} and \eqref{defmu2} imply that $0\leq\ln\mu_+\leq\pi$ and $\pi\leq\ln\mu_-\leq2\pi$.
Consequently, the angle $\tilde\Theta$ defined in Eq.~\eqref{deftheta2} covers the range
\begin{align}
    \Arccos\left(\frac{M}{Q}\right) \leq \tilde\Theta \leq \frac{\sqrt{Q^2-M^2}}{\alpha}\pi + \Arccos\left(\frac{M}{Q}\right). \label{tildetheta}
\end{align}
If, and only if, this range does not contain $\tilde\Theta=k\pi$ for any integer $k$, no singularities arise at least in terms of $f$.
This requirement introduces an additional regularity condition.
Specifically, $\tilde\Theta$ does not encounter any singular points if and only if
\begin{align}
    0<\Arccos\left(\frac{M}{Q}\right) < \frac{\alpha-\sqrt{Q^2-M^2}}{\alpha}\pi \label{minimum}
\end{align}
holds\footnote{We assumed both $M$ and $Q$ are positive, which implies that $0<\Arccos(M/Q)<\pi/2$.}.

The regularity condition on the throat \eqref{regularity1} determines the amount of scalar charge $\Q$ from other parameters $M$, $Q\,(=\sqrt{Q_\mathrm{e}^2+Q_\mathrm{m}^2})$ and $\alpha$.
Consequently, the inequality~\eqref{minimum} implies that there is a lower bound on phantom charge, \textit{i.e.}, negative energy, to construct an entirely regular solution.
To see this, we rewrite the right hand side of Eq.~\eqref{minimum} as
\begin{align}
    \frac{\alpha-\sqrt{Q^2-M^2}}{\alpha}\pi = \left(1-\sqrt{B-\frac{\Q^2}{\alpha^2}}\right)\pi.
\end{align}
The regularity condition on the throat~\eqref{sheetnumber1} requires $B\geq1$.
Therefore, the scalar charge $\Q$ must be sufficiently large to simultaneously satisfy two independent regularity conditions, imposed on the throat~\eqref{regularity1} and at \ms infinity~\eqref{minimum}.

We also comment on the mass observed at \ms infinity, $M_{(-)}$.
An asymptotic analysis similar to that in Sec.~\ref{asymptotic} reveals that the mass is identified with $M_{(-)}=-Q\cos\tilde\Theta_{-\infty}$, where
\begin{align}
    \tilde\Theta_{-\infty} &\coloneqq \frac{\sqrt{Q^2-M^2}}{\alpha}\pi + \Arccos\left(\frac{M}{Q}\right).
\end{align}
The angle $\tilde\Theta_{-\infty}$ corresponds to the maximum value of $\tilde\Theta$ in Eq.~\eqref{tildetheta}, and the regularity condition~\eqref{minimum} prevents it from reaching $\tilde\Theta_{-\infty}=\pi$.
This gives an upper bound on the mass, $M_{(-)}<+Q$.

\subsection{``Extremal'' case} \label{extremal}
The argument $\Theta$ defined in Eq.~\eqref{deftheta1} seems to be singular in the limit $Q\to M$, but the combination $\sqrt{M^2-Q^2}/\sinh\Theta$ remains regular.
Indeed, in the limit $Q\to M$, $\Theta$ behaves as
\begin{align}
    \Theta &\approx \frac{\sqrt{M^2-Q^2}}{M}\left(1+\frac{M}{2\alpha}\ln\mu\right).
\end{align}
Then, one can find
\begin{align}
    f &= \frac{1}{\left(1+\frac{M}{2\alpha}\ln\mu\right)^2}
\end{align}
The corresponding Maxwell potential, with $Q_\mathrm{m}=0$ for simplicity, is given by
\begin{align}
    A_\mu\d x^\mu = \frac{1}{1+\frac{M}{2\alpha}\ln\mu}\d t,
\end{align}
up to a constant.
The scalar field $\Psi$ remains unchanged 
under taking the limit,
\begin{align}
    \Psi &= \frac{P}{2\alpha}\ln\mu.
\end{align}
Interestingly, the functions $f$, $A_t$ and $\Psi$ can be expressed using a single function,
\begin{align}
    H \coloneqq 1 + \frac{M}{2\alpha}\ln\mu,
\end{align}
where $H$ is a harmonic function in three-dimensional Euclidean space $\mathbb{R}^3$, namely, $D^2H=0$ holds.
After choosing appropriate gauges, $f$, $A_t$ and $\Psi$ are written as
\begin{align}
    f &= \frac{1}{H^2}, \qquad A_t = \frac{1}{H}, \qquad \Psi = \frac{\Q}{M}H. \label{harmonics}
\end{align}
Conversely, if one supposes the forms~\eqref{harmonics} and the staticity, $\omega=0$, the Ernst equations~\eqref{ernsteq} are satisfied if the common function $H$ satisfies the Laplace equation, $D^2H=0$.
This result implies that one can superpose any number of extremal wormholes, forming multi-centered solutions.

Next, let us consider the case with magnetic charge, \textit{i.e.}, $Q_\mathrm{m}\neq0$.
While $A_\phi$ cannot be explicitly expressed in terms of $H$, 
the twisted potential $\tilde A_\phi$, given as the imaginary part of the complex electromagnetic potential $\Phi$ in Eq.~\eqref{Phipot5}, can be.
Referring to Eq.~\eqref{whpot1}, we have
\begin{align}
    A_t = \frac{Q_\mathrm{e}}{M}\frac{1}{H}, \qquad \tilde A_\phi = \frac{Q_\mathrm{m}}{M}\frac{1}{H},
\end{align}
where $Q_\mathrm{e}^2+Q_\mathrm{m}^2=M^2$.

\section{Spheroidal coordinates} \label{spheroid}
We have presented the charged solutions in the Weyl-Papapetrou form, namely, in $(\rho, z)$ coordinates.
While this coordinate system is convenient for generating new solutions, its functional form is often complicated.
Since the two-sheet solution, $n = 1$, is expected to have spherical symmetry, 
there should exist a simple coordinate system that reflects the symmetry.
Indeed, this is the case. 
Furthermore, even for any multi-sheet solution, 
similar spheroidal coordinates, which cover one complete sheet and adjacent ones, can be introduced to simplify the metric form. 
In this section, we provide a coordinate transformation to obtain these simplified spheroidal coordinates, which will be useful for analyzing the physical properties of the solutions.

\begin{figure}[t]
    \centering
    \includegraphics{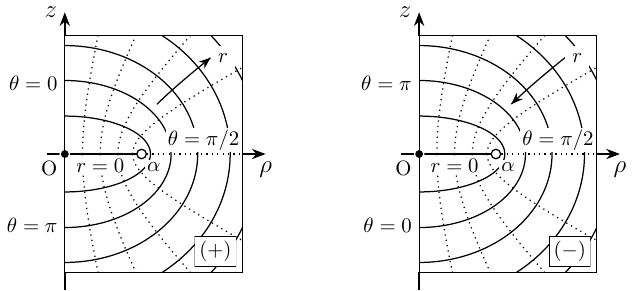}
    \caption{
    The contour curves of $(r, \theta)$ coordinates introduced in the individual $\rho z$ planes.
    The left panel describes the \ps sheet, while the right panel shows the \ms sheet.
    The thick curves denote the contours of $r$ while the dotted denote those of $\theta$.
    These coordinates approach to the usual spherical coordinates in the limit $r\to+\infty$.
    In our multi-sheet wormhole case, the contours of $r$ coincide with those of the solitons $\mu_\pm$, and hence, they also represent the contours of $f$, $\Psi$ and $A_t$ (in addition, the radial magnetic field, $B_r$).
    Note that this coordinate system is known as the oblate spheroidal coordinates, and the disk at $r=0 \ (0\leq\rho<\alpha, z=0)$ is a coordinate singularity where the polar angle $\theta$ becomes multi-valued.
    We remove it by gluing two types of sheets.
    }
    \label{figspheroid}
\end{figure}

\begin{figure}[t]
    \centering
    \includegraphics{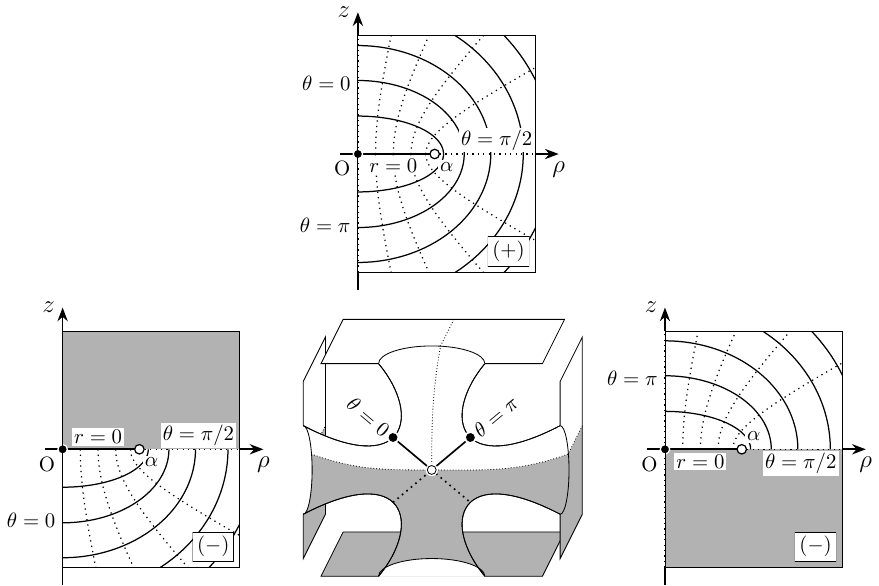}
    \caption{
    An example of covering by single spheroidal coordinates for $n=2$ wormhole spacetimes.
    A pair of spheroidal coordinates covers one complete region in one \ps sheet and two half regions in neighboring \ms sheets.
    For $n=1$, corresponding to a standard two-sided wormhole, the two planes whose half regions are covered coincide;
    the spheroidal coordinates cover the entire spacetime.
    For $n\ge2$, in contrast, trajectories along constant-$\theta$ curves terminate in different asymptotic regions according to whether $\theta<\pi/2$ or $\theta>\pi/2$.
    Our verification presented in Sec.~\ref{edge} confirmed the regularity of the edge denoted by $\circ$, $(r, \theta)=(0, \pi/2)$; it is just a coordinate singularity.
    }
    \label{figspheroid2}
\end{figure}

We apply the coordinate transformation,
\begin{align}
    \left\{\
    \begin{aligned}
        \rho &= \sqrt{r^2+\alpha^2}\sin\theta, \\
        z &= r\cos\theta,
    \end{aligned}
     \right. \label{coord1}
\end{align}
where we assume that the \textit{radial} coordinate $r$ introduced in \ps sheet is non-negative ($r\geq0$) while that in \ms sheet is assumed to be non-positive ($r\leq0$).
These two independent coordinate patches meet at the segment of the throat, $r=0$ (see Figs.~\ref{figspheroid} and \ref{figspheroid2}).
Under this transformation, $w$ defined in Eq.~\eqref{defw} becomes
\begin{align}
    w(r) &= \frac{r^2-\alpha^2}{r^2+\alpha^2}
\end{align}
and the solitons $\mu_\pm$ are able to be rewritten as a unified form,
\begin{align}
    \mu(r) &\coloneqq \left\{\,
    \begin{aligned}
        \mu_+(r) 
        \quad & &\text{for} \quad r\geq0, \\
        \mu_-(r) 
        \quad & &\text{for} \quad r\leq0,
    \end{aligned}
    \right.
    \\
    &= \exp\left[2\Arccos\left(\frac{r}{\sqrt{r^2+\alpha^2}}\right)\right],
\end{align}
which makes it possible to extend the coordinate of a \ps sheet to the adjacent \ms sheets (see Fig.~\ref{figspheroid2}).

Using this expression, one obtains a preferable form of the solutions, \textit{i.e.}, the solution in spheroidal coordinates.
As an example, we first show the results for the uncharged solution presented in Sec.~\ref{prelim}.
Following the coordinate transformation~\eqref{coord1}, we obtain
\begin{align}
    \d \bar{s}^2 &= -\bar{f}(r)\d t^2 + \frac{1}{\bar{f}(r)}\left\{\left(\frac{r^2+\alpha^2}{r^2+\alpha^2\cos^2\theta}\right)^{B-1}\left[\d r^2+\left(r^2+\alpha^2\right)\d\theta^2\right]+\left(r^2+\alpha^2\right)\sin^2\theta\d\phi^2\right\} \label{spheroidal1}
\end{align}
and the corresponding scalar field,
\begin{align}
    \bar{\Psi}(r) &= \frac{\Q}{\alpha}\Arccos{\left(\frac{r}{\sqrt{r^2+\alpha^2}}\right)}, \label{spheroidal4}
\end{align}
where $\bar f(r)$ is given in Eq.~\eqref{deff1}, rewritten in terms of $r$ as
\begin{align}
    \bar f(r) &= \exp\left[-\frac{2m}{\alpha}\Arccos\left(\frac{r}{\sqrt{r^2+\alpha^2}}\right)\right],
\end{align}
and the radial coordinate $r$ takes any real value.
One can easily verify the ADM mass observed at $r\to +\infty$ is $m$, and that at $r\to -\infty$ is $-me^{m\pi/\alpha}$, as shown in Ref.~\cite{Makita:2025bao}.

This reformulation can be carried over to the charged case without any modifications.
Following the same coordinate transformation, one finds
\begin{align}
    \d s^2 &= -f(r)\d t^2 + \frac{1}{f(r)}\left\{\left(\frac{r^2+\alpha^2}{r^2+\alpha^2\cos^2\theta}\right)^{B-1}\left[\d r^2+\left(r^2+\alpha^2\right)\d\theta^2\right]+\left(r^2+\alpha^2\right)\sin^2\theta\d\phi^2\right\}, \label{spheroidal2}
\end{align}
where $f(r)$ is given in Eq.~\eqref{deff3} together with Eq.~\eqref{deftheta1} as
\begin{align}
    f(r) &= \frac{M^2-Q^2}{Q^2}\frac{1}{\sinh^2\Theta(r)}, \\
    \Theta(r) &= \frac{\sqrt{M^2-Q^2}}{\alpha}\Arccos\left(\frac{r}{\sqrt{r^2+\alpha^2}}\right) + \ln\left(\frac{M+\sqrt{M^2-Q^2}}{Q}\right).
\end{align}
The corresponding Maxwell field can also be written in $(r, \theta)$ coordinates.
It is simple to express it in a gauge invariant form, namely the field strength tensor given in Eq.~\eqref{faraday}.
We find
\begin{align}
    \frac{1}{2}F_{\mu\nu}\d x^\mu\wedge\d x^\nu &= -\frac{Q_\mathrm{e}}{r^2+\alpha^2}f(r)\d t\wedge\d r + Q_\mathrm{m}\sin\theta\d\theta\wedge\d\phi.
    \label{spheroidal3}
\end{align}
This expression indicates that the strength of the electromagnetic field depends only on the radial coordinate $r$.
The only non-vanishing component of the electric field, $E_r$, is given by~\cite{Misner:1973prb, Carroll:2004st}
\begin{align}
    E_r &\coloneqq F_{rt} = \frac{Q_\mathrm{e}}{r^2+\alpha^2}f(r),
\end{align}
while that of the magnetic field, $B_r$, can be expressed as
\begin{align}
    B_r &\coloneqq \frac{F_{\theta\phi}}{\sqrt{g_{\theta\theta}g_{\phi\phi}}} = \frac{Q_\mathrm{m}}{r^2+\alpha^2}f(r).
\end{align}
From these expressions, it is obvious that the electromagnetic field has a radial profile.

It is worth mentioning that the asymptotic behaviors of $f(r)$ in the limits $r\to\pm\infty$.
In the positive limit $r\to+\infty$, it admits the following expansion in terms of $1/r$:
\begin{align}
    f &= 1 - \frac{2M}{r} + \O{\frac{1}{r^2}}.
\end{align}
Namely, the radial coordinate $r$ approaches to $\rp$ in this limit.

Similarly, in the negative limit $r \to -\infty$, it behaves as
\begin{align}
    f &= \frac{M^2-Q^2}{Q^2}\frac{1}{\sinh^2\X} \left[1+\frac{2\sqrt{M^2-Q^2}\coth\X}{\abs{r}}\right] + \O{\frac{1}{\abs{r}^2}},
\end{align}
where the quantity $\X$ defined in Eq.~\eqref{defX} is introduced again.
The asymptotic relations between the current coordinates $(t, r, \theta)$ and the coordinate normalized at spatial infinity of \ms sheet $(t_{(-)}, \rm, \thm)$ are given by
\begin{align}
    t &\approx \left(\frac{Q}{\sqrt{M^2-Q^2}}\sinh\X\right)t_{(-)}, \qquad r \approx -\left(\frac{\sqrt{M^2-Q^2}}{Q}\frac{1}{\sinh\X}\right)\rm, \qquad \theta = \pi-\thm.
\end{align}
To see the electromagnetic charge observed at this spatial infinity, we express the Maxwell field in these spheroidal coordinates.
Referring to Sec.~\ref{asymptotic}, we evaluate them in the limit $r\to-\infty$ as
\begin{align}
    A_\mu\d x^\mu &\approx -\frac{M^2-Q^2}{Q^2}\frac{1}{\sinh^2\X}\frac{Q_\mathrm{e}}{\abs{r}}\d t - Q_\mathrm{m}\cos\theta\d\phi \\
    &\approx +\frac{Q_\mathrm{e}}{\rm}\d t_{(-)} + Q_\mathrm{m}\cos\thm\d\phi, \\
    \frac{1}{2}F_{\mu\nu}\d x^\mu\wedge\d x^\nu &\approx +\frac{Q_\mathrm{e}}{\rm^2}\d t_{(-)}\wedge\d\rm - Q_\mathrm{m}\sin\thm\d\thm\wedge\d\phi
\end{align}
One concludes that the ADM mass~$M_{(-)}$, the electric charge~$Q_{\mathrm{e}(-)}$ and the magnetic charge~$Q_{\mathrm{m}(-)}$ observed at this spatial infinity are respectively identified with $-Q\cosh\X$, $-Q_\mathrm{e}$ and $-Q_\mathrm{m}$.
While we have shown the results for $M^2>Q^2$ here, those for $M^2\leq Q^2$ can be obtained in an analogous manner.

\section{Summary and Discussion} \label{summary}
We construct charged wormhole solutions in four-dimensional Einstein-Maxwell-massless phantom scalar system with Lorentzian signature.
The wormhole structure is supported by a phantom scalar field and a monopole-like electromagnetic field.
We apply the Harrison transformation, which generates a charged solution from a neutral seed.
Our seed solution is the neutral multi-sheet wormhole solution found in Ref.~\cite{Makita:2025bao} that connects an even number of asymptotic regions.
The obtained solutions are composed of $2n$ sheets, half of which are called \ps sheets and the others are \ms in this paper.
The configurations of the metric, the Maxwell field and the scalar field in each sheet are summarized in Eqs.~\eqref{charged1a}-\eqref{charged1c} in the cylindrical coordinates and Eqs.~\eqref{spheroidal4}, \eqref{spheroidal2}, and \eqref{spheroidal3} in the spheroidal coordinates.
These solutions are characterized by five physical parameters:
the mass $M$, the electric charge $Q_\mathrm{e}$, the magnetic charge $Q_\mathrm{m}$, the scalar charge $\Q$, and the throat radius $\alpha$.
We identify the ADM mass observed at \ps infinity~$M_{(+)}$ with $M$, and the conserved charges~$Q_{\mathrm{e}{(+)}}$ and $Q_{\mathrm{m}(+)}$ at that infinity are $Q_\mathrm{e}$ and $Q_\mathrm{m}$.
The regularity condition imposes a constraint among these parameters given in Eq.~\eqref{regularity1}.
Consequently, only four of them, as well as the number of sheets $2n$, remain independent parameters of the solution.

As shown in Sec.~\ref{asymptotic}, the electromagnetic charges in the \ms sheets have signs opposite to those in the \ps sheets, $Q_{\mathrm{e}(+)}=-Q_{\mathrm{e}(-)}$ and $Q_{\mathrm{m}(+)}=-Q_{\mathrm{m}(-)}$.
The absolute values of the ADM masses generally differ between the \ps sheet and the \ms sheet,
$M_{(-)}<-M_{(+)}$ for usual ``sub-extremal'' cases.
This discrepancy arises because the normalization of the time coordinate in the \ms sheets differs from that in the \ps sheets. 
Note that, by the construction of the solution, there are no electromagnetic sources present anywhere in the spacetime.
The boundaries of this spacetime consist only of the asymptotic infinities;
the spacetime is entirely free of singularities in contrast to charged black holes.
 
It is important to emphasize that our charged solutions admit an extension beyond mass-charge balancing limit, $M^2=Q^2$.
The solutions we derived above by the Harrison transformation describe ``sub-extremal'' situations, $M^2>Q^2$.
In this case, the mass observed at the \ms infinity must be negative, $M_{(-)} < -M_{(+)}$.
Then we extend the solutions to ``overcharged'' situations, $M^2<Q^2$, by simply replacing $\sqrt{M^2-Q^2}$ with $\I\sqrt{Q^2-M^2}$.
In this case, an additional constraint~\eqref{minimum} is required to avoid a singularity arising in the \ms sheet.
Contrary to the ``sub-extremal'' case, the mass in the \ms sheet can be positive;
$M_{(-)}$ lies in the range from $-M_{(+)}$ to $+Q$.
The regularity condition imposed at the throat, Eq.~\eqref{regularity1}, shows that increasing the electromagnetic charge reduces the amount of negative energy required to support the throat.
However, this does not mean that it is possible to construct a wormhole without a phantom field ($\Psi=0$). 
This can be seen from the fact that, in cases with excessively large electromagnetic charges, two different regularity conditions described in Eqs.~\eqref{regularity1} and \eqref{minimum} can never be satisfied simultaneously.
This is consistent with the fact that the wormhole spacetimes generally require the violation of energy conditions at least on the throat~\cite{Morris:1988tu, MattVisser, Hochberg:1997wp}.

Let us give some remarks on the ``extremal'' limit, $M^2=Q^2$.
In this case, the mass in the opposite side $M_{(-)}$ coincides with $-M_{(+)}$.
Such symmetry appears to be part of the enhanced symmetries specific to the extremal limit.
It is worth considering a generalization to multi-centered wormholes, which connect different asymptotic infinities via multi throats.
In this case, as mentioned in Sec.~\ref{extension}, the metric function $f$, the electrostatic potential $A_t$ (together with the magnetostatic dual potential $\tilde A_\phi$) and the scalar field $\Psi$ can be written in terms of a single harmonic function $H$, which satisfies the Laplace equation in three-dimensional Euclidean space $\mathbb{R}^3$ with axial symmetry.
Actually, we can superpose the solutions in a manner analogous to the multi-centered extremal black hole solutions, known as the Majumdar-Papapetrou solutions.
The primary differences will appear in the regularity conditions.
In the ``extremal'' situation $M^2=Q^2$, the attractive gravitational force and the repulsive electrostatic force are balanced. 
Due to this balance, static multi-black hole solutions can be maintained.
In multi-wormhole cases, however, the additional phantom scalar field $\Psi$ may contribute to an additional repulsive force, which breaks this balance.
As a result, cosmic strings would generically appear between wormholes to provide the necessary tension to maintain a static configuration. 
A simple superposition may produce such cosmic strings, which geometrically manifest as conical deficits. 
These defects could potentially be removed by introducing an additional force field, such as an external electromagnetic field, as demonstrated by Ernst for the $C$-metric~\cite{Ernst:1976mzr} and Emparan for the magnetic dipole~\cite{Emparan:1999au}.
We leave this problem for future work.

\section*{Acknowledgment}
This work was financially supported by JST SPRING, Grant Number JPMJSP2125.
The author Y.~M. would like to take this opportunity to thank the ``THERS Make New Standards Program for the Next Generation Researchers.''
K.~I. and D.~Y. are supported by Japan Society for the Promotion of Science (JSPS), JP24K07046(K.~I.), and 26K07084 (D.~Y.).

\appendix

\section{Detailed regularity analyses} \label{analyses}
We provide here the details of analyses performed in Sec.~\ref{regularities}.
The procedures follow the same approach as those given in Ref.~\cite{Makita:2025bao}.

\subsection{Regularity at the edge of the cut}
\label{app: edge}
In Sec.~\ref{edge}, we introduce the polar coordinate~\eqref{polar1} around the edge $(\rho, z)=(\alpha, 0)$ and expand the functions as series of $\tilde r$.
Being aware of the ambiguity of the square root functions, 
where the branch cut of the square root is chosen along the negative real axis,
the resulting expressions for the solitons $\mu_{\pm\I\alpha}$ introduced in Eq.~\eqref{soliton2} are given as
\begin{align}
    \mu_{\pm\I\alpha} &= \pm\I\alpha \pm\I\sqrt{2\alpha}\,\e^{\mp\I(n\tilde\theta)}\,\tilde r^n + \sin\,(2n\tilde\theta)\,\tilde r^{2n} + \O{\tilde r^{3n}} \label{sum2p}
\end{align}
for the \ps sheets, while
\begin{align}
    \mu_{\pm\I\alpha} &= \pm\I\alpha \mp\I\sqrt{2\alpha}\,\e^{\mp\I(n\tilde\theta)}\,\tilde r^n + \sin\,(2n\tilde\theta)\,\tilde r^{2n} + \O{\tilde r^{3n}} \label{sum2m}
\end{align}
for the \ms sheets.
Referring to the expressions of $\ln\mu_\pm$ in Eq.~\eqref{soliton4}, one finds
\begin{align}
    \ln\mu &= \pi - 2\sqrt{2}\,\sin\,(n\tilde\theta)\,\left(\frac{\tilde r^n}{\sqrt{\alpha}}\right) + \O{\left(\frac{\tilde r^n}{\sqrt{\alpha}}\right)^{3}}. \label{sum1}
\end{align}
Note that the branch cut of the logarithm should be chosen along the negative real axis.
From this result, one can find the fact that they satisfy the Laplace equation, $D^2\ln\mu$, and that the overall factor in Eq.~\eqref{conical} converges to a non-zero finite value.

Likewise above, the magnetic potential $A_\phi$ can be expressed in series of $\tilde r$.
From Eq.~\eqref{sum2p} and \eqref{sum2m}, we find
\begin{align}
    \mu_{-\I\alpha}-\mu_{+\I\alpha} &= -2\I\alpha - 2\I\sqrt{2\alpha}\cos\,(n\tilde\theta)\,\tilde r^n + \O{\tilde r^{3n}}
\end{align}
for the \ps sheets, and
\begin{align}
    \mu_{+\I\alpha}-\mu_{-\I\alpha} &= +2\I\alpha - 2\I\sqrt{2\alpha}\cos\,(n\tilde\theta)\,\tilde r^n + \O{\tilde r^{3n}}
\end{align}
for the \ms sheets.
From Eq.~\eqref{magnetic1}, we obtain
\begin{align}
    A_\phi^\pm &= \pm Q_\mathrm{m}+\sqrt{2}\,Q_\mathrm{m}\cos\,(n\tilde\theta)\,\frac{\tilde r^n}{\sqrt{\alpha}} + \O{\left(\frac{\tilde r^n}{\sqrt{\alpha}}\right)^3}. \label{A phi in r}
\end{align}
Although the constant term has a different sign in each sheet, it can be gauged away.
Ignoring this constant, $A_\phi$ is written as 
\begin{align}
    A_\phi &=  \sqrt{2}\,Q_\mathrm{m}\cos\,(n\tilde\theta)\,\frac{\tilde r^n}{\sqrt{\alpha}} + \O{\left(\frac{\tilde r^n}{\sqrt{\alpha}}\right)^3}.
\end{align}
This ensures that $A_\phi$ is regular at the edge, $\tilde r =0$.

\subsection{Regularity on the axis}
\label{app: axis}
In Sec.~\ref{axis}, we expand the metric functions as series in $\rho$ in the vicinity of the axis of rotation.
The expanded form of the metric function~$f_\pm$ can be obtained as follows:
Firstly, referring to the results in Ref.~\cite{Makita:2025bao}, we have
\begin{align}
    w &= \frac{z^2-\alpha^2}{z^2+\alpha^2} + \O{\rho^2}, \\
    \ln\mu_+ &= \underbrace{\Arccos\left(\frac{z^2-\alpha^2}{z^2+\alpha^2}\right)}_{\eqqcolon \, \ln\mu_0^+} + \O{\rho^2}, \\
    \ln\mu_- &= \underbrace{2\pi-\Arccos\left(\frac{z^2-\alpha^2}{z^2+\alpha^2}\right)}_{\eqqcolon \, \ln\mu_0^-} + \O{\rho^2}. 
    \end{align}
Substituting the above into Eqs.~\eqref{deftheta1} and~\eqref{deff3}, we obtain
\begin{align}
    \Theta_\pm &= \underbrace{\frac{\sqrt{M^2-Q^2}}{2\alpha}\ln\mu_0^\pm + \ln\left(\frac{M+\sqrt{M^2-Q^2}}{Q}\right)}_{\eqqcolon \, \Theta_0^\pm} + \O{\rho^2} \\
    f_\pm &= \underbrace{\frac{M^2-Q^2}{Q^2}\frac{1}{\sinh^2\Theta_0^\pm}}_{\eqqcolon \,f_0^\pm}+\O{\rho^2}. 
\end{align}
Note that $f_0^\pm$ is a constant, inverse of which is the one appearing as $(\textrm{constant})$ in Eq.~\eqref{axisreg}. 
Consequently, One find that the regularity condition to avoid a conical singularity on the axis is given as $A=1$.

\subsection{Asymptotic flatness} \label{app: asym}
In Sec.~\ref{asymptotic}, we verify the asymptotic behavior in each sheet. This is performed by introducing an expansion parameter
\begin{align}
    \epsilon \coloneqq \frac{\alpha}{\sqrt{\rho^2+z^2}}.
\end{align}
In the \ps sheet, we can expand the line element in series of $\epsilon$ as 
\begin{align}
    \d s^2 &= -\left(1-\frac{2M}{\alpha}\epsilon+\O{\epsilon^2}\right)\d t^2 - \left(1-\frac{2M}{\alpha}\epsilon+\O{\epsilon^2}\right)^{-1}\left\{A\left(1+\O{\epsilon^4}\right)\left(\d\rho^2+\d z^2\right)+\rho^2\d\phi^2\right\}
\end{align}
and the Maxwell field, up to a constant, as
\begin{align}
    A_\mu^+\d x^\mu &= \left(-\frac{Q_\mathrm{e}}{\alpha}\epsilon + \O{\epsilon^2}\right)\d t + \left(-\frac{Q_\mathrm{m}}{\alpha}z\epsilon + \O{\epsilon^3}\right)\d\phi.
\end{align}
By introducing the normalized coordinates given in Eq.~\eqref{polar2p}, the line element~\eqref{charged1a} and the Maxwell field~\eqref{charged1b} in asymptotic \ps region become, respectively
\begin{align}
    \d s_+^2 &\approx -\left(1-\frac{2M}{\rp}\right)\d t^2 + \left(1+\frac{2M}{\rp}\right)\left(A\d\rp^2 + A\rp^2\d\theta^2 + \rp^2\sin^2\theta\d\phi^2\right), \label{params3}
\end{align}
and, up to a constant,
\begin{align}
    A_\mu^+\d x^\mu &\approx -\frac{Q_\mathrm{e}}{\rp}\d t - Q_\mathrm{m}\cos\theta\d\phi, \\
    \frac{1}{2}F_{\mu\nu}^+\d x^\mu\wedge\d x^\nu &\approx -\frac{Q_\mathrm{e}}{\rp^2}\d t\wedge\d\rp + Q_\mathrm{m}\sin\theta\d\theta\wedge\d\phi, \label{params4}
\end{align}
respectively.
These expressions state that the solution is asymptotically flat if and only if $A=1$ holds.
Under this condition, we can identify the ADM mass of this system~$M_{(+)}$ evaluated at the \ps infinity with $M$, while the electric charge~$Q_{\mathrm{e}(+)}$ and magnetic charge~$Q_{\mathrm{m}(+)}$ at that infinity are $Q_\mathrm{e}$ and $Q_\mathrm{m}$, respectively.

A similar analysis can be performed for the \ms sheet.
Using the rescaled time $t_{(-)}$ and spherical coordinates $(\rm, \thm)$ given in Eq.~\eqref{polar2m}, 
the expansion parameter is expressed as
\begin{align}
    \epsilon &= \left(\frac{Q}{\sqrt{M^2-Q^2}}\sinh\X\right)\frac{\alpha}{\rm} + \O{\frac{1}{\rm^2}}.
\end{align}
Then, the asymptotic forms of the line element and the gauge field are obtained as
\begin{align}
    \d s_-^2 &\approx -\left(1 + \frac{2Q\cosh\X}{\rm}\right)\d t_{(-)}^2 + \left(1 - \frac{2Q\cosh\X}{\rm}\right)\left(A\d\rm^2 + A\rm^2\d\thm^2 + \rm^2\sin^2\thm\d\phi^2\right) \label{params5}
\end{align}
and, up to a constant,
\begin{align}
    A_\mu^-\d x^\mu &\approx +\frac{Q_\mathrm{e}}{\rm}\d t_{(-)} + Q_\mathrm{m}\cos\thm\d\phi, \\
    \frac{1}{2}F_{\mu\nu}^-\d x^\mu\wedge\d x^\nu &\approx +\frac{Q_\mathrm{e}}{\rm^2}\d t_{(-)}\wedge\d\rm - Q_\mathrm{m}\sin\thm\d\thm\wedge\d\phi, \label{params6}
\end{align}
where the quantity $\X$ is defined in Eq.~\eqref{defX}.
As in the previous case, the condition $A=1$ ensures asymptotic flatness in the \ms sheet as well.
The physical quantities at \ms infinity, ADM mass~$M_{(-)}$, the electric charge~$Q_{\mathrm{e}(-)}$ and the magnetic charge~$Q_{\mathrm{m}(-)}$, are identified with $-Q\cosh\X,\, -Q_\mathrm{e}$ and $-Q_\mathrm{m}$, respectively.

We can verify that the mass $M_{(-)}=-Q\cosh\X$ converges to $-M\e^{M\pi/\alpha}$ in the limit $Q\to 0$, which is consistent with the results for the neutral wormhole analysed in Ref.~\cite{Makita:2025bao}. 
In the ``extremal'' limit $Q\to M$, $M_{(-)}=-Q\cosh\X$ converges to $-M$, matching the result obtained for the exactly ``extremal'' case discussed in Sec.~\ref{extremal}.
We also mention that the mass observed in the \ms infinity can be positive if the electromagnetic charges are sufficiently larger than the mass, as given in Sec.~\ref{overcharged}.

\bibliography{reference.bib}
\bibliographystyle{JHEP}

\end{document}